\documentclass[10pt, letterpaper]{article}
\usepackage{setspace}
\usepackage{fullpage} %or \usepackage[margin=1.0in]{geometry}
\usepackage{hyperref}
\usepackage{graphicx}
\usepackage{xspace}

\newcommand{\shortonly}[1]{}

\newcommand{\lncsonly}[1]{}
\newcommand{\articleonly}[1]{#1}
\newcommand{\acmonly}[1]{}
\newcommand{\svonly}[1]{}

\newcommand{\myparagraph}[1]{\paragraph{#1.}}
\newcommand{\mycomment}[1]{}

\acmonly{
\copyrightyear{2017} 
\acmYear{2017} 
\setcopyright{acmcopyright}
\acmConference{SACMAT '17}{June 21--23, 2017}{Indianapolis, IN, USA}
\acmPrice{15.00}
\acmDOI{http://dx.doi.org/10.1145/3078861.3078878}
\acmISBN{978-1-4503-4702-0/17/06}
}

\articleonly{}
\lncsonly{}
\acmonly{}

\newcommand{\tuple}[1]{\langle #1\rangle}
\newcommand{\mean}[1]{\left[ \! \left[ #1 \right]\! \right]}
\newcommand{\setc}[2]{\{#1 \;|\; #2\}}
\newcommand{\intersect}{\cap}
\newcommand{\ourlang}{ORAL\xspace}
\newcommand{\cm}{{\it CM}}
\newcommand{\om}{{\it OM}}
\newcommand{\type}{{\rm type}}
\newcommand{\stype}{{\rm sType}}
\newcommand{\scond}{{\rm sCond}}
\newcommand{\rtype}{{\rm rType}}
\newcommand{\rcond}{{\rm rCond}}
\newcommand{\con}{{\rm con}}
\newcommand{\acts}{{\rm acts}}
\newcommand{\Act}{{\it Act}}
\newcommand{\nav}{{\rm nav}}
\newcommand{\val}{{\it val}}
\newcommand{\values}{{\it vals}}
\newcommand{\op}{{\it op}}
\newcommand{\Rules}{{\it Rules}}
\newcommand{\spa}{{\it SP}}
\newcommand{\computecond}{{\rm computeCondition}}
\newcommand{\ind}{\hspace*{0.7em}}
\newcommand{\graph}{{\rm graph}}
\newcommand{\reach}{{\rm reach}}
\newcommand{\paths}{{\rm paths}}
\newcommand{\rmpath}{{\rm rm}}
\newcommand{\spath}{{\rm sPath}}
\newcommand{\rpath}{{\rm rPath}}
\newcommand{\Qsp}{Q_{\rm sp}}
\newcommand{\Qac}{Q_{\rm ac}}
\newcommand{\isid}{{\rm isId}}

\newcommand{\mspl}{{\rm MSPL}}
\newcommand{\mrpl}{{\rm MRPL}}
\newcommand{\mtpl}{{\rm MTPL}}
\newcommand{\sped}{{\rm SPED}}
\newcommand{\rped}{{\rm RPED}}
\newcommand{\mcse}{{\rm MCSE}}

\newcommand{\union}{\cup}

\newcommand{\set}[1]{\{#1\}}

\newcommand{\uncovsp}{{\it uncovSP}}

\newcommand{\cc}{{\it cc}}
\newcommand{\gen}{{\it gen}}
\newcommand{\subcond}{{\it subCond}}
\newcommand{\rescond}{{\it resCond}}
\newcommand{\subtype}{{\it subType}}
\newcommand{\restype}{{\it resType}}
\newcommand{\constr}{{\it constr}}
\newcommand{\actions}{{\it acts}}
\newcommand{\results}{{\it results}}
\newcommand{\isempty}{{\rm isEmpty}}
\newcommand{\add}{{\rm add}}

\newcommand{\removeall}{{\rm removeAll}}

\newcommand{\Qpol}{Q_{\rm pol}}
\newcommand{\toString}{\rm toString}
\newcounter{lnum}

\newcommand{\ssub}{s_{\rm s}}
\newcommand{\sa}{s_{\rm a}}
\newcommand{\rhobest}{\rho_{\rm best}}
\newcommand{\rhomerge}{\rho_{\rm mrg}}

\newcommand{\remove}{{\it remove}}

\newcommand{\freq}{{\rm freq}}

\newcommand{\valid}[1]{{\rm valid}(#1)}
\newcommand{\com}[1]{{\it #1}}
\newcommand{\function}{{\bf function}}
\newcommand{\ifstmt}{{\bf if}}
\newcommand{\elsestmt}{{\bf else}}
\newcommand{\eifstmt}{{\bf end if}}
\newcommand{\forloop}{{\bf for}}
\newcommand{\foreachloop}{{\bf for each}}
\newcommand{\whileloop}{{\bf while}}
\newcommand{\eforloop}{{\bf end for}}
\newcommand{\ewhileloop}{{\bf end while}}
\newcommand{\breakstmt}{{\bf break}}

\newcommand{\forloopin}{{\bf in}}
\newcommand{\forloopto}{{\bf to}}
\newcommand{\return}{{\bf return}}

\newcommand{\Copy}{{\rm copy}}
\newcommand{\addcandidaterule}{{\rm addCandidateRule}}
\newcommand{\mul}{{\rm multiplicity}}
\newcommand{\opfrommul}{{\rm opFromMul}}
\newcommand{\candidateconstraint}{{\rm candidateConstraint}}

\newcommand{\generalizerule}{{\rm generalizeRule}}

\newcommand{\mergeandsimplify}{{\rm mergeRulesAndSimplify}}  % formerly mergeRulesLUBandSimplify
\newcommand{\mergerules}{{\rm mergeRules}}  % formerly mergeRulesLUB
\newcommand{\mergerulesinher}{{\rm mergeRulesInheritance}}
\newcommand{\Qrul}{Q_{\rm rul}}

\newcommand{\simplifyrules}{{\rm simplifyRules}}
\newcommand{\wsc}{{\rm WSC}}
\newcommand{\far}{{\rm FAR}}
\newcommand{\frr}{{\rm FRR}}
\newcommand{\ID}{{\rm ID}}

\newcommand{\rmpathL}{\hyperlink{rmpath}{\rmpath}}
\newcommand{\mulL}{\hyperlink{mul}{\mul}}
\newcommand{\navL}{\hyperlink{nav}{\nav}}
\newcommand{\opfrommulL}{\hyperlink{opfrommul}{\opfrommul}}
\newcommand{\reachL}{\hyperlink{reach}{\reach}}
\newcommand{\pathsL}{\hyperlink{paths}{\paths}}

\newcommand{\addcandidateruleL}{\hyperlink{addcandidaterule}{\addcandidaterule}}
\newcommand{\computecondL}{\hyperlink{computecond}{\computecond}}
\newcommand{\candidateconstraintL}{\hyperlink{candidateconstraint}{\candidateconstraint}}

\newcommand{\generalizeruleL}{\hyperlink{generalizerule}{\generalizerule}}

\newcommand{\mergeandsimplifyL}{\hyperlink{mergeandsimplify}{\mergeandsimplify}}\newcommand{\mergerulesinherL}{\hyperlink{mergerulesinher}{\mergerulesinher}}
\newcommand{\mergerulesL}{\hyperlink{mergerules}{\mergerules}}
\newcommand{\QspL}{\hyperlink{Qsp}{\Qsp}}
\newcommand{\QrulL}{\hyperlink{Qrul}{\Qrul}}

\newcommand{\simplifyrulesL}{\hyperlink{simplifyrules}{\simplifyrules}}

\newcommand{\wscRuleL}{\hyperlink{wscRule}{\wsc}}
\begin{document}

\newcommand{\thanksText}{This material is based on work supported in part by %
    NSF Grants %
    CNS-1421893, % trustworthy policies
    and CCF-1414078, % clarity for distrib
    ONR Grant N00014-15-1-2208, % diversity
    AFOSR Grant FA9550-14-1-0261, % ARRIVE
    and DARPA Contract FA8650-15-C-7561. % MARPLE
    Any opinions, findings, and conclusions or recommendations expressed in
    this material are those of the authors and do not necessarily reflect
    the views of these agencies.}

\title{Greedy and Evolutionary Algorithms for Mining Relationship-Based Access Control Policies\lncsonly{\thanks{\thanksText}}\articleonly{\thanks{\thanksText}}}
% do not use \titlenote when using \thanks
%\acmonly{\titlenote{Produces the permission block, and copyright information}}

\articleonly{\author{Thang Bui and Scott~D.~Stoller and Jiajie Li\\
    Department of Computer Science, Stony Brook University, USA}}

\lncsonly{\author{Thang Bui \and Scott~D.~Stoller \and Jiajie Li}
  \institute{Department of Computer Science, Stony Brook University, USA}}

\acmonly{
\author{Thang Bui}
\affiliation{
\institution{Stony Brook University}}
% sigconf format displays all of the following info
%\department{Department of Computer Science}
%\streetaddress{New Computer Science, MSC 2424}
%\city{Stony Brook}
%\state{NY}
%\postcode{11794-2424}
%\country{U.S.A.}
\email{thang.bui@stonybrook.edu}
\author{Scott~D.~Stoller}
\affiliation{
\institution{Stony Brook University}}
\email{stoller@cs.stonybrook.edu}
\author{Jiajie Li}
\affiliation{
\institution{Stony Brook University}}
\email{jiajie.li@stonybrook.edu}
}

\svonly{\author{Thang Bui \and Scott~D.~Stoller \and Jiajie Li}

\institute{T. Bui \and S. D. Stoller \and J. Li \at Stony Brook University, USA\\ \email{stoller@cs.stonybrook.edu}}

\date{Received: date / Accepted: date}}

% acm: abstract must precede maketitle
% lncs (and probably other styles): abstract must follow maketitle.
\newcommand{\abstracttext}{ %
  Relationship-based access control (ReBAC) provides a high level of expressiveness and flexibility that promotes security and information sharing.  We formulate ReBAC as an object-oriented extension of attribute-based access control (ABAC) in which relationships are expressed using fields that refer to other objects, and path expressions are used to follow chains of relationships between objects.

ReBAC policy mining algorithms have potential to significantly reduce the cost of migration from legacy access control systems to ReBAC, by partially automating the development of a ReBAC policy from an existing access control policy and attribute data.  This paper presents two algorithms for mining ReBAC policies from access control lists (ACLs) and attribute data represented as an object model: a greedy algorithm guided by heuristics, and a grammar-based evolutionary algorithm.  An evaluation of the algorithms on four sample policies and two large case studies demonstrates their effectiveness.
}

% The algorithms can be adapted to mine ReBAC policies from access logs and object models. 
%They are the first algorithms for these problems.

\acmonly{
\begin{abstract}
\abstracttext
\end{abstract}}

% from http://www.sheridanprinting.com/typedept/sacmat.htm#L
% Authors please note that the chairs wish to collect the ACM 2012 Classifiers (CCS Concepts) related to your submission. With this step on the submission page, you will NOT have to include the ACM 2012 Classifiers (CCS Concepts) section on the first page of your submission. 

% the following is acmonly but cannot be used in a macro argument
% The code below should be generated by the tool at http://dl.acm.org/ccs.cfm
% <concept_id>10002978.10002991.10002993</concept_id>
% <concept_desc>Security and privacy~Access control</concept_desc>
% \ccsdesc[500]{Security and privacy~Access control}

%\acmonly{\keywords{security policy mining; attribute-based access control; relationship-based access control}}

\acmonly{\thanks{\thanksText}}

\maketitle

\lncsonly{
\begin{abstract}
\abstracttext
\end{abstract}}
\articleonly{
\begin{abstract}
\abstracttext
\end{abstract}}
\svonly{
\begin{abstract}
\abstracttext
\end{abstract}}

% actually, ``Computers & Security submission only''
\articleonly{\noindent {\bf Keywords}: access control policy mining; relationship-based access control; attribute-based access control; evolutionary algorithms; access control policy development}

% !TeX root = main.tex

\section{Introduction}
\label{sec:intro}

The term {\it relationship-based access control} (ReBAC) was introduced to describe access control policies expressed in terms of interpersonal relationships in social network systems (SNSs) \cite{gates2007access}.  The underlying principle of expressing access control policies in terms of chains of relationships between entities is equally applicable and beneficial in general computing systems: it increases expressiveness and often allows more natural policies.  This paper presents \ourlang (Object-oriented Relationship-based Access-control Language), a ReBAC language formulated as an object-oriented extension of ABAC.  Relationships are expressed using attributes that refer to other objects, including subjects and resources, and path expressions are used to follow chains of relationships between objects.  In \ourlang, a ReBAC policy consists of a class model, an object model, and access control rules.  Section \ref{sec:related} compares \ourlang with previous ReBAC models.

High-level access control policy models such as ABAC and ReBAC are becoming increasingly important, as policies become more dynamic and more complex.  This is reflected in the widespread transition from access control lists (ACLs) to role-based access control (RBAC), and more recently in the ongoing transition from ACLs and RBAC to attribute-based access control (ABAC).  In industry, more and more products support ABAC, using a standardized ABAC language such as XACML \cite{XACML} or a vendor-specific ABAC language.  In government, the Federal Chief Information Officer Council called out ABAC as a recommended access control model \cite{FEDCIO11,NIST13ABAC-URL}.  ABAC allows ``an unprecedented amount of flexibility and security while promoting information sharing between diverse and often disparate organizations'' \cite{NIST13ABAC-URL}.  ABAC and ReBAC overcome some of the problems associated with RBAC, notably role explosion \cite{NIST13ABAC-URL}, which makes RBAC policies large and hard to manage.  High-level policy models allow concise policies and promise long-term cost savings through reduced management effort.

The cost of manually developing an initial high-level policy is a barrier to adoption of high-level policy models \cite{NIST13ABAC-URL}.  {\em Policy mining} algorithms promise to drastically reduce this cost, by partially automating the process.
% short: There is a significant amount of research on role mining and some recent research on ABAC policy mining \cite{xu15miningABAC,xu13miningABACfromRBAC,xu14miningABAClogs,medvet2015}.
Role mining, i.e., mining of RBAC policies, is an active research area 
% omitted: zhang07role,ma10role,takabi10stateminer,vaidya08migrating
% omitted: colantonio08cost-driven,conf/sacmat/FrankBB10,frank08class,molloy09evaluating,
% omitted: molloy10noisy,molloy12generative,lu12constraint,zhang13evolving,
(e.g., \cite{conf/sacmat/kuhlm03,conf/sacmat/schle05,vaidya06roleminer,vaidya07finding,guo08role,fuchs2008hydro,lu08optimal,colantonio09formal,molloy10mining,vaidya10role,colantonio12decomposition,verde12role,xu12algorithms,frank13role,xu13mining,mitra16trbac,mitra2016survey,stoller17mining}) and a relatively small (about \$70 million as of 2012) but rapidly growing commercial market segment \cite{hachana12role}.  Role mining is supported by several commercial products, including CA Technologies Identity Governance, Courion RoleCourier, IBM Tivoli Access Manager, Oracle Identity Analytics, NEXIS contROLE, and Novell Access Governance Suite.  Research on ABAC policy mining is in the early stages, with initial work by Xu and Stoller \cite{xu15miningABAC,xu13miningABACfromRBAC,xu14miningABAClogs} and Medvet, Bartoli, Carminati, and Ferrari \cite{medvet2015}.  There is no prior work on mining of ReBAC policies (or object-oriented ABAC policies with path expressions).

This paper defines the ReBAC policy mining problem and presents the first algorithms for mining ReBAC policies from ACLs and attribute data represented as object models.  It is easy to show that the problem is NP-hard, based on Xu and Stoller's proof that ABAC policy mining is NP-hard \cite{xu15miningABAC}.  Since we desire an efficient and practical algorithm, our algorithms are not guaranteed to generate an optimal policy.

Our {\em greedy algorithm}, based on Xu and Stoller's algorithm for mining ABAC policies from ACLs \cite{xu15miningABAC}, has three phases.  In the first phase, it iterates over tuples in the subject-permission relation, uses selected tuples as seeds for constructing candidate rules, and attempts to generalize each candidate rule to cover additional tuples in the subject-permission relation by replacing conditions on user attributes or resource attributes with constraints that relate user attributes with resource attributes.  The algorithm greedily selects the highest-quality generalization according to a rule quality metric based primarily on the ratio of the number of previously uncovered subject-permission tuples covered by the rule to the size of the rule.  The first phase ends when the set of candidate rules covers the entire subject-permission relation.  The second phase attempts to improve the policy by merging and simplifying candidate rules.  The third phase selects the highest-quality candidate rules for inclusion in the mined policy.

Our {\em evolutionary algorithm}, inspired by Medvet et al.'s evolutionary algorithm for mining ABAC policies, uses grammar-based genetic programming \cite{whigham1995,gramEvolSurvey2010}.  It has two phases.  In the first phase, it iterates over tuples in the subject-permission relation, and uses each of the selected tuples as the seed for an evolutionary search that adds one new rule to the candidate policy.
%; it repeats this process until the entire subject-permission relation is covered.  
Each evolutionary search starts with an initial population containing candidate rules created from a seed tuple in a similar way as in our greedy algorithm along with numerous random variants of those rules together with some completely random candidate rules, evolves the population by repeatedly applying genetic operators (mutations and crossover), and then selects the highest quality rule in the population as the result of that evolutionary search.  The second phase attempts to improve the candidate rules by further mutating them.

% Our algorithms can be extended to identify suspected noise in the input ACL policy using a similar approach as in \cite{xu15miningABAC}.  Our algorithms can be adapted to mine ReBAC policies from access logs and object models, in a similar way as Xu and Stoller's algorithm \cite{xu15miningABAC} was adapted to mine ABAC policies from access logs and attribute data \cite{xu14miningABAClogs}.

We evaluate our algorithms on four relatively small but non-trivial sample policies and two larger and more complex case studies, based on Software-as-a-Service (SaaS) applications offered by real companies \cite{decat14edoc,decat14workforce}. To the best of our knowledge, the latter are the largest rule-based policies (as measured by the number and complexity of the rules) used in the evaluation of any policy mining algorithm targeting a rule-based policy language.

Our evaluation methodology is to start with a ReBAC policy, generate ACLs representing the subject-permission relation, run a policy mining algorithm on the generated ACLs (along with the class model and object model), and compare the ReBAC policy mined from ACLs with the original ReBAC policy.  For the four sample policies, both of our policy mining algorithms achieve optimal or nearly optimal results.  For the case studies, our greedy algorithm and evolutionary algorithm achieve 84\% and 91\% (respectively) average syntactic similarity between the mined policy and a simplified but equivalent version of the original policy.  Experiments on object models of varying size for the two case studies show that both algorithms have good performance and scale reasonably well: as a function of the number of subject-permission tuples, the running time of the greedy algorithm is less than quadratic, and the running time of the evolutionary algorithm is close to linear.
% and have similar performance on the workforce management case study, and the evolutionary algorithm is slower on the e-document case study.

We conclude that both algorithms produce high-quality mined policies that, if used as a starting point for development for a ReBAC policy, would save the policy developers a significant amount of effort.

This paper is a revised and greatly extended version of a SACMAT 2017 short paper \cite{bui17mining}, which briefly described an earlier version of our greedy algorithm and presented experimental results for it.  The most significant addition in this paper is our evolutionary algorithm, and the experimental results comparing the effectiveness and performance of our two algorithms.  Other notable additions are more detailed description of our greedy algorithm,
% through addition of more pseudocode (Figures \ref{fig:computecond} and \ref{fig:generalizerule}) and more explanatory text, 
an example that illustrates the working of our greedy algorithm,
% (Section \ref{sec:algorithm:example}), 
and more detailed descriptions of the sample policies and case studies.
% (Section \ref{sec:sample-policies}),
%and updated and expanded experimental results.
% (Section \ref{sec:eval}).
% and expanded discussion of related work (Section \ref{sec:related}).

% not a main addition: performance evaluation using more policies,

% !TeX root = main.tex

\section{Related Work}
\label{sec:related}

We discuss related work on policy models and related work on policy mining.
\subsection{Policy Models}

Entity-Based Access Control (EBAC) \cite{bogaerts15entity} is the policy model most closely related to ours.  EBAC is quite similar to \ourlang, except that it is based on entity-relationship models, instead of object-oriented models, and hence lacks the concept of inheritance, which \ourlang includes.  EBAC's expression language includes quantifiers, and \ourlang does not, although some conditions that require quantifiers in their language can be expressed in \ourlang using the built-in binary relations on sets, such as $\supseteq$.

Several ReBAC models have been proposed, by Carminati, Ferrari, and Perego \cite{carminati2009enforcing}, Fong \cite{fong2011rebac-model}, Cheng, Park, and Sandhu \cite{cheng2012user}, Hu, Ahn, and Jorgensen \cite{hu2013multiparty}, Crampton and Sellwood \cite{crampton2014path}, and others.  Some are designed specifically for OSNs, while others are designed for general use.  Our model differs from all of them because it is designed as a (nearly) minimal extension of a typical ABAC language, and the extension is achieved by adopting an object-oriented model and incorporating standard object-oriented concepts, notably path expressions, like in UML's Object Constraint Language (OCL) (\url{http://www.omg.org/spec/OCL/}).  None of these ReBAC models are based on general object-oriented data models.  None of these ReBAC models can express constraints relating fields (a.k.a. attributes) of different entities, such as the constraint subject.affiliation $\in$ resource.patient.registrations in the ORAL rule in Equation (\ref{eq:emr-rule}).  In this regard, \ourlang is significantly more expressive.

%\ourlang includes inheritance, although it is not essential for ReBAC, because it allows more natural and concise domain models.

On the other hand, \ourlang lacks some features found in some of these ReBAC models, such as transitive closure (e.g., supervisor$^*$ refers to the subject's supervisor, the subject's supervisor's supervisor, and so on), negation (e.g., dept $\ne$ CS), and graph patterns (which can specify more than a single path).  Many realistic applications do not require these language features; on the other hand, they are useful for some applications.  These features can easily be added to our policy language.  However, developing policy mining algorithms that fully exploit them may be difficult.  That challenge is future work.

%longer version of the above
%On the other hand, \ourlang lacks some features found in some of these ReBAC models.  For example, all of the languages cited above include some form of transitive closure (for example, supervisor$^*$ may refer to the subject's supervisor, the subject's supervisor's supervisor, and so on), and \ourlang does not.  The languages in \cite{fong2011rebac-model,cheng2012user,huth2012hybrid,bogaerts15entity} include some form of negation (for example, dept $\ne$ CS), and \ourlang does not.  The modal-logic-based policy languages in \cite{fong2011rebac-model,huth2012hybrid} include formulas that specify graph patterns, not merely paths.  Many realistic applications do not require these language features; on the other hand, they are useful for some applications.  These features can easily be added to our policy language.  However, developing policy mining algorithms that fully exploit them may be difficult.  We leave that challenge for future work.

%The languages in \cite{fong2011rebac-model,huth2012hybrid,bogaerts15entity} include some form of quantifiers, and \ourlang does not, although some conditions expressed with quantifiers in other frameworks can be expressed in \ourlang using atomic constraints of the form $\tuple{p_1, {\rm supseteq}, p_2}$. 

The languages in \cite{fong2011rebac-model,cheng2012user,huth2012hybrid,crampton2014path} allow every relation to be traversed in reverse.  \ourlang, like EBAC and OCL, does not; instead, the policy designer explicitly enables reverse traversal where appropriate by including a field in the reverse direction (this corresponds to using a bidirectional association in the UML class model).

Our access control model can be characterized as {\em object-oriented ABAC}. We prefer to characterize it as ReBAC to emphasize the difference from typical ABAC languages such as XACML.  In XACML, attributes values are primitive values, such as numbers or strings, or collections thereof, not object references.  Primitive values can be (and often are) object identifiers, but they cannot be dereferenced, so policies that require path expressions cannot be expressed.  For example, XACML can express the condition user.dept=CS but not user.dept.college=ENG.  This limitation is typically circumvented by duplicating information, e.g., introducing an attribute user.college.  This workaround is inefficient and increases the administrative burden, because user.college must be updated whenever user.dept is updated; in our framework, user.dept.college automatically has the correct value after user.dept is updated.  Next Generation Access Control (NGAC) is an ABAC standard being developed at NIST \cite{NIST16comparison-short}.  Its data model is richer than XACML's, allowing nested collections of entities, but it does not adopt a general object-oriented view in which subjects, resources, and other types of objects are modeled in a uniform way, and it does not support general path expressions.

%There are some ABAC models for object-oriented databases and object-oriented programming languages, e.g., \cite{essmayr97}.  They are specialized to fine-grained context-sensitive control of field access and method invocation.  Our model is more generic.

%Jeffrey Fischer, Daniel Marino, Rupak Majumdar, and Todd Millstein
%Fine-Grained Access Control with Object-Sensitive Roles, ECOOP 2009
%http://web.cs.ucla.edu/~todd/research/ecoop09.pdf
%based on RBAC, not ABAC

\subsection{Policy Mining}

There is no prior work on mining of ReBAC policies (or object-oriented ABAC policies with path expressions).  The most closely related prior work on policy mining is for ABAC policies without path expressions.

Xu and Stoller developed the first algorithms for mining ABAC policies, from attribute data plus ACLs \cite{xu15miningABAC}, roles \cite{xu13miningABACfromRBAC}, or access logs \cite{xu14miningABAClogs}.  Our greedy algorithm is based on their algorithm for mining ABAC policies from ACLs \cite{xu15miningABAC}.  Adapting their algorithm to be suitable for ReBAC mining required many changes, most notably generalization of loops over attributes to iterate over paths when generating conditions and constraints; specifically, we introduce the idea of generating constraints based on paths between classes in the graph representation of the class model.  The technique for merging rules for sibling classes into a rule for an ancestor class is also new.  We also modified the algorithm to accommodate changes in the supported relational operators: in conditions, we allow ``in'' and ``contains'', instead of ``equal'' and ``supseteq'' in \cite{xu15miningABAC}; in constraints, we allow ``in'' in addition to ``equal'', ``contains'', and ``supseteq'' allowed in \cite{xu15miningABAC}. Other algorithm differences include deferring removal of redundant rules (by modifying $\mergerulesL$ not to remove redundant rules, and removing redundant rules before final rule selection) and adding a third component to the rule quality metric.  We also introduce several techniques to limit and prioritize the paths being considered, since naively considering all type-correct paths would make the algorithm prohibitively expensive, even for small policies.  For example, when generating constraints, we base them only on the shortest and nearly-shortest paths between classes in the class model.

%Prioritization includes the specifying order in which constraints are considered in $\generalizeruleL$ (the example in Section \ref{sec:algorithm} illustrates the importance of this) and the order in which conditions are considered in $\simplifyrulesL$; these processing orders are unspecified and uncontrolled in \cite{xu15miningABAC}, because in the simpler context of ABAC mining, they did not affect the results.}

%Limits include type constraints, limits on length of individual paths, and limits on the total path length in a constraint.  

% also, generalization of the definitions of policy quality and rule quality, but this is straightforward.

Medvet {\it et al.}'s evolutionary algorithm for ABAC policy mining \cite{medvet2015} inspired our evolutionary algorithm for ReBAC policy mining.  Our algorithm, like theirs, has an evolutionary search phase using the separate-and-conquer strategy, followed by an improvement phase.  The separate-and-conquer strategy \cite{bartoli15learning}, which in the context of policy mining means learning one rule at a time, instead of an entire policy at once, is essential to obtain good results.  We also adopt their fitness function, which, in turn, is based on Xu and Stoller's rule quality metric \cite{xu15miningABAC}.  A key difference is that Medvet et al.'s algorithm uses an ad-hoc application-specific genotype (i.e., representation of individuals) together with genetic operators specifically designed to operate on that genotype.  In contrast, we adopt the general and well-studied framework of grammar-based genetic programming \cite{whigham1995, gramEvolSurvey2010}: we represent individuals as derivation trees, and we use genetic operators that operate on derivation trees.  We use the classical mutation and crossover operators on derivation trees but also introduce a few more genetic operators specialized to the general structure of our grammars (e.g., non-terminals for conditions and actions are treated differently), which enable the evolutionary search to produce high-quality rules much more quickly.  The operators are not specific to details of the predicate language, so our algorithm can easily be applied to extensions of the policy language with additional datatypes and relational operators.  Another difference from Medvet et al.'s algorithm is that our algorithm uses a more complicated construction for the initial population of each evolutionary search.  These features enable our algorithm to achieve good results with reasonable computation time, despite the significantly larger search space for ReBAC policies compared with ABAC policies.  Also, Medvet et al.'s algorithm was evaluated only on policies comparable in size to our four sample policies, not on large policies comparable to our two case studies.

%Evaluation on some hand-written benchmark ABAC policies from \cite{xu15miningABAC} demonstrates that Medvet et al.'s algorithm is effective at minimizing WSC.  Minimization of WSC is merely a heuristic.  The gold standard for policy mining is to generate the policy a human administrator would write.  For those benchmark ABAC policies, Xu and Stoller's algorithm exactly reconstructs the original hand-written rules in more cases than Medvet {\it et al.}'s algorithm.

Cotrini et al.'s algorithm for mining ABAC rules from sparse logs \cite{sparselogs2018} is based on APRIORI-SD, a machine-learning algorithm for subgroup discovery.  ``Sparse'' means that only a small fraction of the possible entitlements appear in the log.  Therefore, the algorithm must extrapolate significantly to determine which entitlements not in the log should be granted, and which should be denied.  They formulate a novel heuristic to identify suspected over-permissiveness of ABAC rules. Their algorithm searches for succinct rules that have high confidence and are not overly permissive according to their heuristic.  %Mining ReBAC policies from sparse logs is an interesting direction for future work.

\section{Policy Language}
\label{sec:language}

This section presents our policy language, \ourlang.  It contains common ABAC constructs, similar to those in \cite{xu15miningABAC}, plus path expressions.

A {\em ReBAC policy} is a tuple $\pi=\tuple{\cm, \om, \Act, \Rules}$, where $\cm$ is a class model, $\om$ is an object model, $\Act$ is a set of actions, and $\Rules$ is a set of rules.

A {\em class model} is a set of class declarations.  A {\em class declaration} is a tuple $\tuple{{\it className}, {\it parent}, {\it fields}}$ where {\it parent} is a class name or the empty string (indicating that the class does not have a parent), and {\it fields} is a set of field declarations.  A {\em field declaration} is a tuple $\tuple{{\it fieldName}, {\it type}, {\it multiplicity}}$, where {\it type} is a class name or Boolean, and {\it multiplicity} is optional, one, or many.  The {\em multiplicity} specifies how many values of the specified type may be stored in the field and is ``one'' (also denoted ``1'', meaning exactly one), ``optional'' (also denoted ``?'', meaning zero or one), or ``many'' (also denoted ``*'', meaning any natural number).  Boolean fields always have multiplicity 1.  Every class implicitly contains a field ``id'' with type String.  We keep the language minimal by not allowing user-defined fields with type string and by omitting other base types (e.g., numbers); they could easily be added.  However, their effect can be achieved using a field that refers to an object having the desired string as its id.  Thus, the set of types in a policy contains Boolean, String, and the names of the declared classes.  A {\em reference type} is any class name (used as a type).

An {\em object model} is a set of objects whose types are consistent with the class model and with unique values in the id field.  An {\em object} is a tuple $\tuple{{\it className}, {\it fieldVals}}$, where {\it fieldVals} is a function that maps the names of fields of the specified class, including the id field and inherited fields, to values consistent with the types and multiplicities of the fields.  The value of a field with multiplicity many is a set. The value of a field with multiplicity one or optional is a single value; the special placeholder $\bot$ is used when a field with multiplicity optional lacks an actual value.  
Let $\type(o)$ denote the type of an object $o$.
%For an object $o=\tuple{c, {\it fv}}$, let $\type(o)=c$ and $\fval(o)={\it fv}$.
%In examples, we represent the function {\it fieldVals} by its graph, i.e., a set of input-output pairs.  
% we don't use \fval, because we don't formally define \nav.

% the path in a condition must be non-empty, because the language lacks constant values with reference type

A {\em condition} is a set, interpreted as a conjunction, of atomic conditions.  We often refer to the atomic conditions as conjuncts.
% Informally, an atomic condition is a condition on the value of one field of one object. 
An {\em atomic condition} is a tuple $\tuple{p, \op, \val}$, where $p$ is a non-empty path, $\op$ is an operator, either ``in'' or ``contains'', and $\val$ is a constant (specifically, a Boolean value or string) or a set of constants.  If $\val$ is a single constant, not a set, we say that it is {\em atomic}.  Note that $\val$ cannot equal or contain the placeholder $\bot$.  A {\em path} is a sequence of field names, written with ``.'' as a separator.  For example, if dept and id are field names, then dept.id is a path.  For readability, we usually write conditions with a logic-based syntax, using ``$\in$'' for ``in'' and ``$\ni$'' for ``contains''.  For example, we may write $\tuple{{\rm dept.id}, {\rm in}, \{{\rm CompSci}\}}$ as ${\rm dept.id} \in \{{\rm CompSci}\}$.  We may use ``='' as syntactic sugar for ``in'' when the constant is a singleton set; thus, the previous example may be written as dept.id=CompSci.  A condition may contain multiple atomic conditions on the same path.

% we do not state the well-formedness requirements on conditions here, because they depend on the type relative to which the path is interpreted, which is known only in the context of a rule.

% for conditions, we need to include id field when appropriate, because the constant must be a string or boolean (or set of them), not an object reference.
% for constraints, we don't need to include id field, because we can interpret the constraint as a relation on object references (or entire objects); this is equivalent to the same relation on the value of the objects' id fields, since id's are unique.  I realize that the implementation includes the id field in constraints, but I think we can regard that as an implementation detail, since it is conceptually unnecessary.

A {\em constraint} is a set, interpreted as a conjunction, of atomic constraints. 
 Informally, an atomic constraint expresses a relationship between the requesting subject and the requested resource, by relating the values of paths starting from each of them.  An {\em atomic constraint} is a tuple $\tuple{p_1, \op, p_2}$, where $p_1$ and $p_2$ are paths (possibly the empty sequence), and $\op$ is one of the following four operators: equal, in, contains, supseteq.  The ``contains'' operator is the transpose of the ``in'' operator.  Implicitly, the first path is relative to the requesting subject, and the second path is relative to the requested resource.  The empty path represents the subject or resource itself.  For readability, we usually write constraints with a logic-based syntax, using ``$=$'' for ``equal'' and ``$\supseteq$'' for ``supseteq'', and we prefix the subject path $p_1$ and resource path $p_2$ with ``subject'' and ``resource'', respectively.  For example, $\tuple{{\rm specialties}, {\rm contains}, {\rm topic}}$ may be written as ${\rm subject.specialties} \ni {\rm resource.topic}$. 
% Other relational operators, such as $\subseteq$, could also be added; we omit them for now, since they are not needed for our case studies.

A {\em rule} is a tuple $\langle {\it subjectType}, {\it subjectCondition}, {\it resourceType},{\it resourceCondition},$ ${\it constraint}, {\it actions}\rangle$, where {\it subjectType} and {\it resourceType} are class names, {\it subjectCondition} and {\it resourceCondition} are conditions, {\it constraint} is a constraint, {\it actions} is a set of actions, and the following well-formedness requirements are satisfied.  Implicitly, the paths in {\it subjectCondition} and {\it resourceCondition} are relative to the requesting subject and requested resource, respectively.  The {\em type of a path} $p$ (relative to a specified class), denoted $\type(p)$, is the type of the last field in the path.  The {\em multiplicity of a path} $p$ (relative to a specified class), denoted \hypertarget{mul}{$\mul(p)$}, is one if all fields on the path have multiplicity one, is many if any field on the path has multiplicity many, and is optional otherwise.

\paragraph{Examples.}
We give three example rules here.  As additional examples, all of the sample policies and case studies described in Section \ref{sec:sample-policies} are available on the web, in the ReBAC Miner release at \url{http://www.cs.stonybrook.edu/~stoller/software/}.  In examples, we prefix the path in the subject condition and resource condition with ``subject'' and ``resource'', respectively, for readability.  Our electronic medical records sample policy contains the rule: A physician can create a medical record associated with a consultation if the physician is not a trainee, the consultation is with the physician, and the patient of the consultation is registered at the hospital with which the physician is affiliated.  This is expressed as
\begin{equation}
\label{eq:emr-rule}
\rho = \langle\, 
\begin{array}[t]{@{}l@{}}
  \mbox{Physician}, \mbox{subject.isTrainee=false}, \mbox{Consultation}, \mbox{true},\\ \mbox{subject = resource.physician $\land$ subject.affiliation $\in$ resource.patient.registrations},\\
  \{\mbox{createMedicalRecord}\} \rangle.
\end{array}
\end{equation}
Our healthcare sample policy contains the rule: A doctor can read an item in a HR for a patient treated by one of the teams of which he/she is a member, if the topics of the item are among his/her specialties.  This is expressed as $\langle\,$Doctor, true, HealthRecordItem, true, subject.teams contains resource.record.patient.treatingTeam $\land$ subject.specialties $\supseteq$ resource.topics, \{read\}$\rangle$, where HealthRecordItem.record is the health record containing the HR item.  Our e-document case study involves a large bank whose policy contains the rule: A project member can read all sent documents regarding the project.  This is expressed as $\langle\,$Employee, subject.employer.id = LargeBank, Document, true, subject.workOn.relatedDoc $\ni$ resource, \{read\}$\rangle$, where Employee.workOn is the set of projects the employee is working on, and Project.relatedDoc is the set of documents related to the project.

% this is not a good motivating example for ReBAC: the paths are short.
%Our project management sample policy contains the rule: A contractor working on a project can read and request to work on a non-proprietary task of the project whose required areas of expertise are among his/her areas of expertise.  This is expressed as $\langle$Contractor, true, Task, resource.isProprietary=false, subject.projects $\ni$ resource.project $\land$ subject.expertise $\supseteq$ resource.expertise, \{read, request\}$\rangle$. 

Well-formedness requirements on rules are as follows.  (1) All paths are type-correct, assuming the subject and resource have type {\it subjectType} and {\it resourceType}, respectively.  (2) (a) The two paths in the constraint have the same type, and (b) this type is not String.  Part (a) reflects the assumption that comparing objects of different types is either meaningless or useless (since it would be equivalent to ``false'').  Part (b) prohibits constraints that compare identifiers of objects with different types, which would be meaningless.  It does not reduce the expressiveness of the model, because a constraint violating it, such as ${\rm specialties.id} \ni {\rm topic.id}$, can be written more simply as ${\rm specialties} \ni {\rm topic}$.  (3) The path in the condition does not have reference type.  This reflects the fact that our language does not allow constants with reference type.  (4) In conditions with operator ``in'', the path has multiplicity optional or one, and the value is a set of constants.  This excludes sets of sets from the model.  (5) In conditions with operator ``contains'', the path has multiplicity many, and the value is atomic.  (6) In constraints with operator ``equal'', both paths have multiplicity optional or one.  (7) In constraints with operator ``in'', the first path has multiplicity optional or one, and the second path has multiplicity many.  (8) In constraints with operator ``contains'', the first path has multiplicity many, and the second path has multiplicity optional or one.  (9) In constraints with operator ``supseteq'', both paths have multiplicity many.

Any class can be used as a subject type, resource type, or both.  For example, one rule could allow doctors to read medical records, and another rule could allow department heads to assign doctors to workgroups.

For a rule $\rho=\tuple{st, sc, rt, rc, c, A}$, let $\stype(\rho)=st$, $\scond(\rho)=sc$, $\rtype(\rho)=rt$, $\rcond(\rho)=rc$, $\con(\rho)=c$, and $\acts(\rho)=A$.

A {\em permission} is a pair $\tuple{r,a}$, where $r$ is an object, and $a$ is an action; it represents authorization to perform action $a$ on resource $r$.  A {\em subject-permission} tuple is a tuple $\tuple{s, r, a}$, where $s$ is an object, and $\tuple{r,a}$ is a permission; it means that subject $s$ has permission $\tuple{r,a}$.  A {\em subject-permission relation} is a set of subject-permission tuples.

% below, I write "a set may be obtained if ..." instead of "a set is obtained if ..." because, if a field along the path has value \bot, then \bot is obtained, even if some other field has multiplicity many.

Given a class model, object model, object $o$, and path $p$, let \hypertarget{nav}{$\nav(o,p)$} be the result of navigating (a.k.a. following or dereferencing) path $p$ starting from object $o$.  The class model and object model are implicit arguments to this relation and the following relations.  We elide these arguments, because in our setting, they are unchanging in the context of a given policy.  The result of navigating might be no value, represented by the placeholder $\bot$, an atomic value, or a set.  A set may be obtained if any field along the path (not necessarily the last field) has multiplicity many.  This is like the semantics of path navigation in UML's Object Constraint Language (OCL) (\url{http://www.omg.org/spec/OCL/}).

An object $o$ {\em satisfies} an atomic condition $c=\tuple{p, \op, \val}$, denoted $o\models c$, if $(\op={\rm in} \land \nav(o,p) \in \val) \lor (\op={\rm contains} \land \nav(o,p) \ni \val)$.  The {\em meaning} of a condition $c$ relative to a class $C$, denoted $\mean{c}_C$ is the set of instances of $C$ (in the implicitly given object model) that satisfy $c$.  A condition $c$ {\em characterizes} a set $O$ of objects of class $C$ if $O$ is the meaning of $c$ relative to $C$.

Objects $o_1$ and $o_2$ {\em satisfy} an atomic constraint $c=\tuple{p_1, \op, p_2}$, denoted $\tuple{o_1,o_2} \models c$, if $(\op={\rm equal} \land \nav(o_1,p_1) = \nav(o_2,p_2)) \lor (\op={\rm in} \land \nav(o_1,p_1) \in \nav(o_2,p_2)) \lor (\op={\rm contains} \land \nav(o_1,p_1) \ni \nav(o_2,p_2)) \lor (\op={\rm supseteq} \land \nav(o_1,p_1) \supseteq \nav(o_2,p_2))$.

A subject-permission tuple $\tuple{s, r, a}$ {\em satisfies} a rule $\rho=\langle st, sc, rt,$ $rc, c, A\rangle$, denoted $\tuple{s, r, a} \models \rho$, if
$\type(s)=st \land s\models sc \land \type(r)=rt  \land r\models rc  \land \tuple{s,r}\models c \land a \in A$.

The {\em meaning} of a rule $\rho$, denoted $\mean{\rho}$, is the subject-permission relation it induces, defined as $\mean{\rho} = \setc{\tuple{s,r,a} \in \om\times\om\times\Act}{\tuple{s,r,a}\models \rho}$.

The {\em meaning} of a  ReBAC policy $\pi$, denoted $\mean{\pi}$, is the 
subject-permission relation it induces, defined as the union of the meanings of its rules.

% !TeX root = main.tex

\section{Problem Definition}
\label{sec:problem}

An {\em access control list (ACL) policy} is a tuple $\tuple{\cm, \om, \Act, \spa_0}$, where $\cm$ is a class model, $\om$ is an object model, $\Act$ is a set of actions, and $\spa_0\subseteq \om\times \om \times \Act$ is a subject-permission relation.  Conceptually, $\spa_0$ is the union of the resources' access control lists.

An ReBAC policy $\pi$ is {\em consistent} with an ACL policy $\langle\cm, \om,$ $\Act,$ $\spa_0\rangle$ if they have the same class model, object model, and actions and $\mean{\pi} = \spa_0$.

An ReBAC policy consistent with a given ACL policy can be trivially constructed, by creating a separate rule corresponding to each subject-permission tuple in the ACL policy, using a condition ``id=...'' to identify the relevant subject and resource.  Of course, such a ReBAC policy is as verbose and hard to manage as the original ACL policy.  Therefore, we must decide: among ReBAC policies consistent with a given ACL policy $\pi_0$, which ones are preferable?  We adopt two criteria.

One criterion is that the ``id' field should be avoided when possible, because policies that use this field are (to that extent) identity-based, not attribute-based or relationship-based.  Therefore, our definition of ReBAC policy mining requires that these attributes are used only when necessary, i.e., only when every ReBAC policy consistent with $\pi_0$ contains rules that use them.

The other criterion is to maximize a policy quality metric.  A {\em policy quality metric} is a function $\Qpol$ from ReBAC policies to a totally-ordered set, such as the natural numbers.  The ordering is chosen so that small values indicate high quality; this is natural for metrics based on policy size.  For generality, we parameterize the policy mining problem by the policy quality metric.

The {\em ReBAC policy mining problem} is: given an ACL policy $\pi_0=\langle \cm, \om,$ $\Act, \spa_0\rangle$ and a policy quality metric $\Qpol$, find a set $\Rules$ of rules such that the ReBAC policy $\pi=\tuple{\cm, \om, \Act, \Rules}$ is consistent with $\pi_0$, uses the ``id'' field only when necessary, and has the best quality, according to $\Qpol$, among such policies.

The policy quality metric that our algorithm aims to optimize is {\em weighted structural complexity} (WSC), a generalization of policy size first introduced for RBAC policies \cite{molloy10mining} and later extended to ABAC \cite{xu15miningABAC}.  Minimizing policy size is consistent with prior work on ABAC mining and role mining and with usability studies showing that more concise access control policies are more manageable \cite{beckerle13formal}.  Informally, the WSC of a ReBAC policy is a weighted sum of the numbers of elements of each kind in the policy.  Formally, the WSC of a ReBAC policy $\pi$, denoted $\hypertarget{wscPol}{\wsc(\pi)}$, is the sum of the $\wsc$ of its rules, defined bottom-up as follows.  The $\wsc$ of an atomic condition $\tuple{p, \op, \val}$ is $|p| + |\val|$, where $|p|$ is the length of path $p$, and $|\val|$ is 1 if $\val$ is an atomic value and is the cardinality of $\val$ if $\val$ is a set.  The $\wsc$ of an atomic constraint $\tuple{p_1, \op, p_2}$ is $|p_1|+|p_2|$.  The $\wsc$ of a condition $c$, denoted $\wsc_{\rm cnd}(c)$, is the sum of the $\wsc$ of the constituent atomic conditions.  The $\wsc$ of a constraint $c$, denoted $\wsc_{\rm cns}(c)$, is the sum of the $\wsc$ of the constituent atomic constraints.  The $\wsc$ of a rule is $\hypertarget{wscRule}{\wsc(\tuple{st, sc, rt, rc, c, A})} =
%   \begin{array}[t]{@{}l@{}}
     w_1\wsc_{\rm cnd}(sc) + w_1\wsc_{\rm cnd}(rc) %\smallskip\\
     {} + w_2\wsc_{\rm cns}(c) + w_3|A|,$
     where $|A|$ is the cardinality of set $A$, and the $w_i$ are user-specified weights.

% defined by
% \begin{eqnarray*}
%   \wsc_{\rm cnd}(c) &=& \sum_{\tuple{p, \op, \val} \in c} |p| + |\val|\\
%   \wsc_{\rm cns}(c) &=& \sum_{\tuple{p_1, \op, p_2} \in c} |p_1| + |p_2|\\
%   \hypertarget{wscRule}{\wsc(\tuple{st, sc, rt, rc, c, A})} &=&
%   \begin{array}[t]{@{}l@{}}
%     w_1\wsc_{\rm cnd}(sc) + w_1\wsc_{\rm cnd}(rc)\smallskip\\
%     {} + w_2\wsc_{\rm cns}(c) + w_3|A|
%   \end{array}\\
%   \wsc(\Rules) &=& \sum_{\rho \in \Rules} \wsc(\rho),
% \end{eqnarray*}
% where $|p|$ is the length of path $p$, $|\val|$ is 1 if $\val$ is an atomic value and is the cardinality of $\val$ if $\val$ is a set, $|s|$ is the cardinality of set $s$, and the $w_i$ are user-specified weights.

%%% Local Variables:
%%% mode: latex
%%% TeX-master: "main"
%%% TeX-PDF-mode: t
%%% End:

% !TeX root = main.tex

\section{Greedy Algorithm}
\label{sec:algorithm}

This section presents our greedy algorithm.  It is based on the ABAC policy mining algorithm in \cite{xu15miningABAC}.  The main differences are summarized in Section \ref{sec:related}.

\begin{figure}[tbp]
\begin{tabular}[t]{@{}l@{}}
// \com{Phase 1: Create a set $\Rules$ of candidate rules that covers $\spa_0$.}\\
$\Rules=\emptyset$\\
// \com{$\uncovsp$ contains tuples in $\spa_0$ that are not covered by $\Rules$}\\
% // \com{$\uncovsp$ contains subject-permission tuples}\\
% // \com{in $\spa_0$ that are not covered by $\Rules$}\\
$\uncovsp=\spa_0.\Copy()$\\
\whileloop\ $\neg\uncovsp.\isempty()$\\
\ind // \com{Use highest-quality uncovered tuple as a ``seed'' for rule creation.}\\
\ind   $\tuple{s,r,a}$ = highest-quality tuple in $\uncovsp$ according to $\QspL$\\
\ind   $\cc = \candidateconstraintL(s, r)$\\
\ind   // \com{$\ssub$ contains subjects with permission $\tuple{r,a}$ and that have}\\
\ind   // \com{the same candidate constraint for $r$ as $s$}\\
\ind   $\ssub = 
  \{s'\in \om \;|\; 
  \begin{array}[t]{@{}l@{}}
    \type(s')=\type(s) \land \tuple{s',r,a}\in \spa_0\\
    {} \land \candidateconstraintL(s',r)=\cc\}
\end{array}$\\
\ind $\addcandidateruleL(\type(s), \ssub, \type(r), \set{r}, \cc, \set{a}, \uncovsp, \Rules)$\\
\ind // \com{$\sa$ is set of actions that $s$ can perform on $r$}\\
% I changed $a' \in \Act$ to $a'$; the possible values of a' are clear from $\tuple{s,r,a'}\in \spa_0$.
\ind   $\sa = \setc{a'}{\tuple{s,r,a'}\in \spa_0}$\\
\ind $\addcandidateruleL(\type(s), \set{s}, \type(r), \set{r}, \cc, \sa, \uncovsp, \Rules)$\\
\ewhileloop\\
// \com{Phase 2: Combine rules using least upper bound and inheritance.}\\
// \com{Also, simplify them and remove redundant rules.}\\
%// \com{Merge rules using least upper bound and simplify them}\\
\mergeandsimplifyL(\Rules)\\
%// \com{Merge rules based on inheritance}\\
\mergerulesinherL(\Rules)\\
\mergeandsimplifyL(\Rules)\\
// \com{Remove redundant rules}\\
\whileloop\ $\Rules$ contains rules $\rho$ and $\rho'$ such that $\mean{\rho}\subseteq\mean{\rho'}$\\
\ind $\Rules.\remove(\rho)$\\
\ewhileloop\\
%\removeredundantrulesL(\Rules)\\
// \com{Phase 3: Select high quality rules into $\Rules'$.}\\
$\Rules'$ = $\emptyset$\\
% we didn't define \mean{\Rules'} for a set \Rules' of rules, but the meaning is
% reasonably obvious. 
% we need the "discarding a rule ..." clause because rules that do not
% cover any uncovered UP could have higher quality than rules that do,
% because the latter rules could have more over-assignments.
Repeatedly move highest-quality rule from $\Rules$ to $\Rules'$ until
$\sum_{\rho\in\Rules'}\mean{\rho} \supseteq \spa_0$,\\  
using $\spa_0\setminus\mean{\Rules'}$ as second argument
to $\Qrul$, and discarding a rule if it does not\\
cover any tuples in $\spa_0$ currently uncovered by $\Rules'$.\\
% increase $\mean{\Rules'}\unte$ \\
%\lnum\> $\uncovsp$ = $\spa_0.\Copy()$\\
%\lnum\> \whileloop\ $\neg\uncovsp.\isempty()$\\
%\lnum\>\> $\rhomax =$ a rule in $\Rules$ with maximal quality, i.e.,\\
%\lnum\>\>\ind $\rhomax \in \argmax_{\rho\in\Rules} \QrulL(\rho, \uncovsp)$\\
%\lnum\>\> $\Rules'.\add(\rhomax)$\\
%\lnum\>\> $\Rules.\removeElt(\rhomax)$\\
%\lnum\>\> $\uncovsp.\removeall(\mean{\rhomax})$\\
%\lnum\> \ewhileloop\\
\return\ $\Rules'$
%\\
%// \com{Repeatedly merge rules using least upper bound and}\\
%// \com{simplify them, until this has no effect}\\
%\function\ $\hypertarget{mergeandsimplify}{\mergeandsimplify(\Rules)}$\\
%$\mergerulesL(\Rules)$\\
%\whileloop\ $
%\begin{array}[t]{@{}l@{}}
%  \simplifyrulesL(\Rules) \mathrel{\&\&} \mergerulesL(\Rules)
%\end{array}$\\
%\ind skip\\
%\ewhileloop
\end{tabular}
\caption{Greedy algorithm for ReBAC policy mining.  Inputs: subject-permission relation $\spa_0$, class model $\cm$, and object model $\om$.  Output: set of rules $\Rules'$.  The algorithm also has numerical parameters MCSE, MSPL, MRPL, SPED, RPED, and MTPL that limit the considered rules, as described in the text.  }
\label{fig:alg}
\end{figure}

Top-level pseudocode appears in Figure \ref{fig:alg}.  It reflects the high-level structure described in Section 1. We refer to the tuples selected in the first statement of the first while loop as {\em seeds}.  The top-level pseudocode is explained by embedded comments.  It calls several functions, described next.  For some functions, we give a description in the text and pseudocode; for others, we give only a description, to save space.  Function names hyperlink to pseudocode for the function, if it is included in the paper, otherwise to the description of the function.

The workset $\uncovsp$ in Figure \ref{fig:alg} is a priority queue sorted in descending lexicographic order by the quality $\QspL$ of the subject-permission tuple.  Informally, $\QspL(\tuple{s,r,a})$ is a triple whose first two components are the frequency of permission $\tuple{r,a}$ and subject $s$, respectively, i.e., their numbers of occurrences in $\spa_0$, and whose third component (included as a tie-breaker to ensure a total order) is the string representation of the tuple.
\begin{eqnarray*}
  \freq(\tuple{r,a}) &=& |\setc{\tuple{s',r',a'}\in\spa_0}{r'=r \land a'=a}|\\
  \freq(s) &=& |\setc{\tuple{s',r',a'}\in\spa_0}{s'=s}|\\
  \hypertarget{Qsp}{\Qsp(\tuple{s,r,a})} &=& \tuple{\freq(\tuple{r,a}), \freq(s), \toString(\tuple{{\it s},{\it r},{\it a}})}
\end{eqnarray*}

The function $\candidateconstraint(s, r)$ in Figure \ref{fig:candidateconstraint} returns a set containing all the atomic constraints that hold between resource $r$ and subject $s$ and satisfy path length constraints described below. It first computes a set $\cc$ of candidate constraints using type-correct short paths to each type $T$ reachable from both $\type(s)$ and $\type(r)$ in $\graph(\cm)$, which is defined to be a graph with a vertex for each class, and an edge from class $c_1$ to class $c_2$ if $c_1$ has a field with type $c_2$.  It then selects and returns the candidate constraints satisfied by $\tuple{s,r}$.  This algorithm infers only constraints where the paths have reference types.  It could easily be extended to infer constraints where the paths have type Boolean, but such constraints do not arise in our case studies.  It uses the following auxiliary functions.  The function \hypertarget{reach}{$\reach(T)$} returns the set of classes reachable from $T$ in $\graph(\cm)$, including their superclasses.  The function \hypertarget{paths}{$\paths(T,T',L)$} returns all paths from $T$ to $T'$ in $\graph(\cm)$ whose length is at most $L$ more than the length of the shortest path from $T$ to $T'$.  This reflects our observation that paths in constraints in case studies are the shortest paths between the relevant types or slightly longer.  $\sped$ (mnemonic for ``subject path extra distance'') and $\rped$ (mnemonic for ``resource path extra distance'') are parameters of the algorithm.  We also observe that the subject path and resource path typically do not both have the maximum allowed length in the same constraint, so we introduce a parameter $\mtpl$ (mnemonic for ``maximum total path length'') that limits the sum of the lengths of these paths in a constraint.

The function $\opfrommulL(m,m')$ returns the relational operator suitable for left and right operands with multiplicity $m$ and $m'$, respectively.
\begin{displaymath}
 \hypertarget{opfrommul}{\opfrommul(m,m')} = (\tuple{m,m'}\!=\!\tuple{{\rm many}, {\rm many}} \,?\, {\rm supseteq} : (m\!=\!{\rm many} \,?\, {\rm contains} : (m'\!=\!{\rm many} \,?\, {\rm in} : {\rm equal})))
\end{displaymath}
%defined by the following case statement on $\tuple{m,m'}$, where underscore is a wildcard pattern:
%\begin{center}
%\begin{tabular}{@{}l@{}}
%$|~\tuple{\many, \many} \rightarrow {\rm supseteq}$\\
%$|~\tuple{\many, \_} \rightarrow {\rm contains}$\\
%$|~\tuple{\_\,, \many} \rightarrow {\rm in}$\\
%$|~\tuple{\_\,, \_} \rightarrow {\rm equal}$.
%\end{tabular}
%\end{center}

% note for \paths: length counts only association edges, not parent edges
%note that \mtpl is potentially useful even if \sped=0 and \rped=0.
%In our case studies, \mtpl eliminates a modest number of unused constraints.

\begin{figure}[tbp]
\begin{tabular}{@{}l@{}}
\hypertarget{candidateconstraint}{\function\ $\candidateconstraint(s, r)$}\\
// \com{$cc$ is the set of type-correct candidate constraints}\\
$cc$ = $\emptyset$ \\
\forloop\ $T$ \forloopin\ ($\reachL(\type(s)) \cap \reachL(\type(r))$)\\
\ind // \com{add candidate constraints where the paths have type $T$}\\
\ind \forloop\ $p_1$ \forloopin\ $\pathsL(\type(s), T, \sped)$\\
\ind \ind \ \forloop\ $p_2$ \forloopin\ $\pathsL(\type(r), T, \rped)$ such that $|p_1|+|p_2| \le \mtpl$\\
\ind \ind \ind \ \ $cc$.\add($\tuple{p_1, \opfrommulL(\mulL(p_1), \mulL(p_2)), p_2}$)\\
\ind \ind \ \eforloop\\
\ind \eforloop\\
\eforloop\\
\return\ $\setc{c \in cc}{\tuple{s,r} \models c}$
%// \com{Return relational operator suitable for left operand with  multiplicity $m$ and right operand with}\\
%// \com{ multiplicity $m'$.   Underscore is a wildcard pattern.}\\
%\hypertarget{opfrommul}{\function\ $\opfrommul(m,m')$}\\
%\ind {\bf match} $\tuple{m,m'}$ {\bf with}\\
%\ind \ind $|~\tuple{\many, \many} \rightarrow \return\ {\rm supseteq}$\\
%\ind \ind $|~\tuple{\many, \_} \rightarrow \return\ {\rm contains}$\\
%\ind \ind $|~\tuple{\_\,, \many} \rightarrow \return\ {\rm in}$\\
%\ind \ind $|~\tuple{\_\,, \_} \rightarrow \return\ {\rm equal}$
\end{tabular}
\caption{Compute candidate constraints for subject $s$ and resource $r$}
\label{fig:candidateconstraint}
\end{figure}

The function $\addcandidateruleL(st, \ssub, rt, s_{\rm r}, \cc, \sa, \uncovsp, \Rules)$ in Figure \ref{fig:addcandidaterule} calls $\computecondL$ to compute conditions $sc$ and $rc$ that characterizes $\ssub$ and $s_{\rm r}$, respectively.  $\mspl$ and $\mrpl$ are the maximum path length for paths in the subject condition and resource condition, respectively; they are parameters of the algorithm.  $\addcandidateruleL$ then constructs a rule $\rho=\tuple{st, sc, rt, rc, \emptyset, \sa}$, calls $\generalizeruleL$ to generalize $\rho$ to $\rho'$ and adds $\rho'$ to candidate rule set $\Rules$. The details of the functions called by $\addcandidaterule$ are described next.

\begin{figure}[tbp]
\begin{tabular}{@{}l@{}}
\hypertarget{addcandidaterule}{\function\ $\addcandidaterule(st, \ssub, rt, s_{\rm r}, \cc, \sa, \uncovsp, \Rules)$} \\
// \com{Construct a rule $\rho$ that covers subject-permission tuples $\setc{\tuple{s,r,a}\in\spa_0}{s\in \ssub\land r\in s_{\rm r} \land a\in \sa}$.}\\
$sc = \computecondL(\ssub, st, \mspl)$;\\
$rc = \computecondL(s_{\rm r}, rt, \mrpl)$\\
$\rho = \tuple{st, sc, rt, rc, \emptyset, \sa}$\\
$\rho' = \generalizeruleL(\rho, \cc, \uncovsp, \Rules)$\\
$\Rules.\add(\rho')$\\
$\uncovsp.\removeall(\mean{\rho'})$
\end{tabular}
\caption{Compute a candidate rule and add it to $\Rules$}
\label{fig:addcandidaterule}
\end{figure}

The function $\computecondL(O,C,L)$ in Figure \ref{fig:computecond} computes a condition $C$ that characterizes the set $O$ of objects of type $C$ using paths of length at most $L$.  A path with multiplicity optional or one appears in at most one conjunct, of the form $\tuple{p, {\rm in}, V}$ where $V$ is the collected values of $o.p$ for $o$ in $O$.   A path with multiplicity many may appear in multiple conjuncts, of the form  $\tuple{p, {\rm contains}, \val}$ where $\val$ is in the intersection of the values of $o.p$ for $o$ in $O$.
%For each such path $p$ for which $o.p$ has a value for all $o$ in $O$, if $p$ has multiplicity optional or one, then a conjunct for $p$ is added to $c$, otherwise $p$ has multiplicity many, and a conjunct for $p$ is added for each value in the intersection of the ...
First, paths not containing the {\rm id} field are considered.  If the resulting condition does not characterize $O$, then (by construction) it is an over-approximation, and a conjunct using the ``id'' field is added to ensure that the resulting condition characterizes $O$.  The condition returned by $\computecondL$ might not be minimum-sized among conditions that characterize $O$: possibly some conjuncts can be deleted without changing the condition's meaning.  We defer minimization of the condition until after the call to $\generalizeruleL$ (described below), because minimizing the condition before that would reduce opportunities to find constraints in $\generalizeruleL$.

\begin{figure}[tbp]
\begin{tabular}{@{}l@{}}
\hypertarget{computecond}{\function\ $\computecond(O, C, L)$}\\
// \com{First try to characterize set $O$ without using ``{\rm id}'' field.}\\
$c = \mbox{new Set()}$\\
\foreachloop\ path $p$ s.t. 
  \begin{tabular}[t]{@{}l@{}}
    $\mbox{($p$ is type-correct starting from $C$)}
    \land |p| \le L \land \mbox{($p$ does not contain ``id'')}$
  \end{tabular}\\
\ind  $\values = \setc{\navL(o,p)}{o \in O}$\\
\ind  \ifstmt\ $\bot \not \in \values$\\
\ind \ind  \ifstmt\ $\mulL(p) \in \set{{\rm optional}, {\rm one}}$\\
\ind \ind \ind      $c.\add(\tuple{p, {\rm in}, vals})$\\
\ind \ind   \elsestmt \ind // \com{$\mulL(p) = {\rm many}$}\\
\ind \ind \ind $I = \mbox{intersection of the sets in $\values$}$\\
\ind \ind \ind \forloop\ $\val$ in $I$\\
\ind \ind \ind \ind $c.add(\tuple{p, {\rm contains}, \val})$\\
\ind \ind \ind \eforloop\\
\ind\ind \eifstmt\\
\ind \eifstmt\\
\eforloop\\
\ifstmt\ $\mean{c}_C \ne O$\\
\ind\ind // \com{``{\rm id}'' field is needed to characterize $O$.}\\
\ind\ind $c.\add(\tuple{{\rm id}, {\rm in}, \setc{\navL(o, {\rm id}}{o \in O}})$\\
\eifstmt\\
\return\ $c$
\end{tabular}
\caption{Compute a condition that characterizes set $O$ of objects of type $C$, using paths of length at most $L$. }
\label{fig:computecond}
\end{figure}

A rule $\rho$ is {\em valid}, denoted $\valid{\rho}$, if $\mean{\rho}\subseteq\spa_0$.

% generalizerule considers all subsequences of the candidate constraints, but not all permutations.

The function $\generalizeruleL(\rho, \cc, \uncovsp, \Rules)$ in Figure \ref{fig:generalizerule} attempts to generalize rule $\rho$ by adding some atomic constraints in $\cc$ to $\rho$ and eliminating the conjuncts of the subject condition and resource condition that use the same paths as those constraints.  A rule obtained in this way is called a {\em generalization} of $\rho$.  It is more general in the sense that it refers to relationships instead of specific values.  The meaning of a generalization of $\rho$ is a superset of the meaning of $\rho$.  In more detail, $\generalizeruleL$ tries to generalize $\rho$ using each constraint in $\cc$ separately, discards the invalid generalizations, sorts the valid generalizations in descending order of the number of covered entitlements in $\uncovsp$, recursively tries to further generalize each of them using constraints from $\cc$ that produced valid generalizations later in the sort order, and then returns the highest-quality rule among them (rule quality is defined below); if no generalizations of $\rho$ are valid, it simply returns $\rho$.  When trying to add an atomic constraint $c$ in $\cc$ to a rule $\rho$, $\generalizeruleL$ first tries removing the conjuncts of the subject condition and resource condition that use the same paths as $c$.  If the resulting rule is invalid, it attempts a more conservative generalization by removing only the conjunct in the subject condition that uses the same path as $c$.  If that rule is also invalid, it instead removes only the conjunct in the resource condition that uses the same path as $c$.  If that rule is also invalid, then there is no valid generalization of $\rho$ using $c$.

% removed to shorten the paper, and because it is nitty-gritty detail.  if restoring this, can also restore reference to it in related.tex.
%$\generalizeruleL$ sorts $\cc$ because the order in which candidate constraints are considered can affect the resulting generalized rule.  For example, suppose adding candidate constraint $c_1$ keeps the rule valid and removes a conjunct in the subject condition, and that adding candidate constraint $c_2$ removes a conjunct in the subject condition and a conjunct in the resource condition, and yields a higher-quality resulting rule if the resulting rule is valid. Suppose, further, that adding $c_2$ keeps the rule valid only if $c_1$ has already been added. In this case, the highest-quality generalization will be created only if $c_1$ is considered before $c_2$. The auxiliary function \hypertarget{sortconstraints}{\sortconstraints($cc, \rho, \uncovsp$)} sorts the constraints in $\cc$ and returns the result as a sequence.  Specifically, it temporarily adds each constraint $c$ in $\cc$ to $\rho$ (in the same way as described above) and, if this leads to a valid generalization of $\rho$, computes the number of subject-permission tuples in $\uncovsp$ covered by the resulting rule, and then sorts the constraints in $\cc$ in descending order by these values.  If adding $c$ does not lead to a valid generalization of $\rho$, $c$ is omitted from the result.

% the ordering has additional components; see algorithm.txt for details

% the second component of Qrul is useful in reconstructing the university
% case-studies with synthetic attribute data.

A {\em rule quality metric} is a function $\Qrul(\rho, \spa)$ that maps a rule $\rho$ to a totally-ordered set, with the order chosen such that larger values indicate higher quality.  The second argument $\spa$ is a set of subject-permission tuples.  Based on our primary goal of minimizing the mined policy's WSC, a secondary preference for rules with more atomic constraints, and a tertiary preference for rules with shorter paths in atomic constraints, we define
\begin{displaymath}
  \hypertarget{Qrul}{\Qrul(\rho, \spa)} = \langle|\mean{\rho}\intersect\spa|/\wscRuleL(\rho),\, |\con(\rho)|, 1/{\rm TCPL}(\rho)\rangle
\end{displaymath}
where ${\rm TCPL}(\rho)$ (``total constraint path length'') is the sum of the lengths of the paths used in the atomic constraints of $\rho$.

% the limits L1 and L2 in changes to generalizeRule in algorithm.txt are effectively unused in current experiments; they are set to values that allow all combinations of candidate constraints to be explored.

The preference for more atomic constraints is a heuristic, based on the observation that rules with more atomic constraints tend to be more general than other rules with the same $|\mean{\rho}\intersect\spa|/\wscRuleL(\rho)$ (such rules typically have more conjuncts) and hence lead to lower WSC for the policy.  In $\generalizeruleL$, $\uncovsp$ is the second argument to $\QrulL$, so $\mean{\rho}\intersect\spa$ is the set of subject-permission tuples in $\spa_0$ that are covered by $\rho$ and not covered by existing rules.

The pseudocode for $\generalizeruleL$ in Figure \ref{fig:generalizerule} uses the following auxiliary functions.  $\spath(c)$ and $\rpath(c)$ are the subject path and resource path, respectively, used in atomic constraint $c$. \hypertarget{rmpath}{$\rmpath(c, p)$} is the condition obtained by removing the atomic condition on path $p$ (if any) from condition $c$.  $a[i..]$ denotes the suffix of array $a$ starting at index $i$.  The loop over $i$ in $\generalizeruleL$ considers all possibilities for the first atomic constraint in $\cc$ that gets added to the constraint of $\rho$.  The function calls itself recursively to determine the subsequent atomic constraints in $\cc$ that get added to the constraint.

\begin{figure}[tbp]
\begin{tabular}[t]{@{}l@{}}
 \hypertarget{generalizerule}{\function\ $\generalizerule(\rho, \cc, \uncovsp, \Rules)$}\\
// \com{split $\rho$ into its components, for convenience.}\\
$\tuple{\subtype,\subcond,\restype,\rescond,\constr,\actions} = \rho$\\
// \com{$\rhobest$ is highest-quality generalization of $\rho$}\\
$\rhobest = \rho$\\
// \com{try to create generalizations of $\rho$ using each constraint in $\cc$.  save them in $\results$.}\\
$\results = \mbox{new Vector()}$\\
\forloop\ $c$ \forloopin\ $\cc$\\
\ind// \com{try to generalize $\rho$ by adding constraint $c$ and eliminating the conjuncts for both}\\
\ind// \com{paths used in $c$.}\\
\ind$\rho' = \langle
\begin{array}[t]{@{}l@{}}
  \subtype, \rmpathL(\subcond, \spath(c)),
  \restype,
  \rmpathL(\rescond, \rpath(c)), \constr\union\set{c}, \actions\rangle
\end{array}$\\
\ind// \com{if the paths in $c$ appear in the conditions in $\rho$ and hence have been eliminated in $\rho'$,}\\
\ind// \com{and $\rho'$ is valid, then add $\tuple{c,\rho'}$ to $\results$.}\\
\ind\ifstmt\ $\spath(c)\mbox{ appears in } \subcond \,\land\, \rpath(c)\mbox{ appears in } \rescond \,\land\, \valid{\rho'}$\\
\ind\ind$\results.{\rm add}(\tuple{c,\rho'})$\\
\ind\elsestmt\\
\ind\ind// \com{try to generalize $\rho$ by adding constraint $c$ and eliminating the conjunct for the subject path in $c$.}\\
\ind\ind$\rho' = \langle
\begin{array}[t]{@{}l@{}}
  \subtype, \rmpathL(\subcond, \spath(c)), \restype,
  \rescond, \constr\union\set{c}, \actions\rangle
\end{array}$\\
\ind\ind\ifstmt\ $\spath(c)\mbox{ appears in } \subcond \,\land\, \valid{\rho'}$\\
\ind\ind\ind$\results.{\rm add}(\tuple{c,\rho'})$\\
\ind\ind\elsestmt\\
\ind\ind\ind// \com{try to generalize $\rho$ by adding constraint $c$ and eliminating the conjunct for the resource path in $c$.}\\
\ind\ind\ind$\rho' = \langle
\begin{array}[t]{@{}l@{}}
  \subtype, \subcond, \restype, \rmpathL(\rescond, \rpath(c)),
  \constr\union\set{c}, \actions\rangle
\end{array}$\\
\ind\ind\ind\ifstmt\ $\rpath(c)\mbox{ appears in } \rescond \,\land\, \valid{\rho'}$\\
\ind\ind\ind\ind$\results.{\rm add}(\tuple{c,\rho'})$\\
\ind\ind\ind\eifstmt\\
\ind\ind\eifstmt\\
\ind\eifstmt\\
\eforloop\\
sort $\results$ in descending order by $Q(\tuple{c,\rho'})$ = number of tuples in $\uncovsp$ covered by $\rho'$\\
\cc' = sequence containing the first components of the tuples in $\results$\\
\gen\ = sequence containing the second components of the tuples in $\results$\\
%// \com{``unzip'' $\results$ into lists of constraints and rules in the}\\
%// \com{first and second components of the tuples, respectively.}\\
% alternative notation: Python-like list comprehension.
%\cc' = [$c$ for $\tuple{c, \_}$ in $\results$]\\
%\gen = [$\rho'$ for $\tuple{\_, \rho'}$ in $\results$]\\
\forloop\ $i$ = 1 \forloopto\ $\results$.length\\
\ind// \com{try to further generalize $\gen[i]$}\\
\ind$\rho'' = \generalizeruleL(\gen[i], \cc'[i\!+\!1\,..], \uncovsp, \Rules)$\\
\ind~~ \ifstmt\ $\QrulL(\rho'', \uncovsp) > \QrulL(\begin{array}[t]{@{}l@{}}
  \rhobest, 
  \uncovsp)
\end{array}$\\
\ind\ind$\rhobest = \rho''$\\
\ind\eifstmt\\
\eforloop\\
\return\ $\rhobest$
\end{tabular}
\caption{Generalize rule $\rho$. }
\label{fig:generalizerule}
\end{figure}

The function \hypertarget{mergeandsimplify}{$\mergeandsimplify(\Rules)$} repeatedly calls 
$\mergerulesL$ and $\simplifyrulesL$ until they have no effect.

The function \hypertarget{mergerules}{\mergerules(\Rules)} attempts to improve the quality of $\Rules$ by merging pairs of rules that have the same subject type, resource type, and constraint by taking the least upper bound of their subject conditions, the least upper bound of their resource conditions, and the union of their sets of actions.  The {\em least upper bound} of conditions $c_1$ and $c_2$, denoted $c_1 \sqcup c_2$, is 
\begin{displaymath}
 \begin{array}{@{}l@{}}
 \{\tuple{p, {\rm in}, \val} \;|\; 
  \begin{array}[t]{@{}l@{}}
    (\exists \val_1,\val_2:\; \tuple{p, {\rm in}, \val_1} \in c_1 \land  \tuple{p, {\rm in}, \val_2} \in c_2\\
\ind\ind {} \land \val = \val_1 \union \val_2)\}
  \end{array}\\
{} \union \{\tuple{p, {\rm contains}, \val} \;|\; 
\begin{array}[t]{@{}l@{}}
  \tuple{p, {\rm contains}, \val} \in c_1\\
  {} \land\ \tuple{p, {\rm contains}, \val} \in c_2)\}.
\end{array}
 \end{array}
\end{displaymath}
% note that a path appearing in only one of c_1 and c_2 is discarded, and that a path that appears in "contains" conditions in both c_1 and c_2 is discarded if the contained element is different in c_1 and c_2, and that p cannot appear in conditions with different operators, because the operator must correspond to p's multiplicity.
Note that the meaning of the merged rule $\rhomerge$ is a superset of the meanings of the rules $\rho_1$ and $\rho_2$ being merged.  If the merged rule $\rhomerge$ is valid, then it replaces $\rho_1$ and $\rho_2$ in $\Rules$.  Rule pairs are considered for merging in descending lexicographic order of rule pair quality, where the quality of a rule pair $\tuple{\rho_1,\rho_2}$ is $\tuple{\max(q_1, q_2), \min(q_1,q_2)}$ where $q_i=\QrulL(\rho_i,\spa_0)$. $\mergerulesL(\Rules)$ updates its argument $\Rules$ in place, and it returns a Boolean indicating whether any rules were merged.

The function \hypertarget{simplifyrules}{$\simplifyrules(\Rules)$} attempts to simplify all of the rules in $\Rules$.  It updates its argument $\Rules$ in place, replacing rules in $\Rules$ with simplified versions when simplification succeeds.  It returns a Boolean indicating whether any rules were simplified.  It attempts to simplify each rule in the following ways.
  
(1) It eliminates conjuncts from the subject and resource conditions when this preserves validity. Removing one conjunct might prevent removal of another conjunct, so it searches for a set of removable conjuncts that maximizes the quality of the resulting rule.  To limit the cost, we introduce a parameter $\mcse$ (mnemonic for ``maximum conjuncts to simplify exhaustively'').  If the number of conjuncts is at most $\mcse$, the algorithm tries removing every subset of conjuncts.  If the number of conjuncts exceeds $\mcse$, the algorithm sorts the conjuncts in descending lexicographic order by $\Qac$ (quality metric for atomic conditions) and then attempts to remove them linearly in the sorted order, where $\Qac(\tuple{p, \op, \val}) = \tuple{|\val|, |p|, \isid(p), \toString(p)}$, where $|\val|$ is 1 if $\val$ is atomic and is the cardinality of $\val$ is a set, and $\isid(p)$ is 1 if $p$ is ``id'' and is 0 otherwise.  The last component of $\Qac$ is included as a tie-breaker to ensure a total order.
(2) It eliminates atomic constraints when this preserves validity.  It searches for the set of atomic constraints to remove that maximizes the quality of the resulting rule, while preserving validity.
  (3) It eliminates overlapping actions between rules.  Specifically, an action $a$ in a rule $\rho$ is removed if there is another rule $\rho'$ in the policy such that $\scond(\rho') \subseteq \scond(\rho) \land \rcond(\rho') \subseteq \rcond(\rho) \land \con(\rho') \subseteq \con(\rho) \land a \in \acts(\rho')$.
  (4) It eliminates actions when this preserves the meaning of the policy. In other words, it removes an action $a$ in rule $\rho$ if all the subject-permission tuples covered by $a$ in $\rho$ are covered by other rules in the policy.  Note that (3) is a special case of (4), listed separately to ensure that this special case takes precedence.
(5) If the subject condition contains an atomic condition of the form $p=c$, and the constraint contains an atomic constraint of the form $p=p'$, then replace that atomic constraint with the atomic condition $p'=c$ in the resource condition (note that this is a form of constant propagation); and similarly for the symmetric situation in which the resource condition contains such an atomic condition, etc.
%
% removing the cycle might change the multiplicity of the path.  in this case, we change the relational operator as necessary, and then check validity of the resulting rule.
(6) Remove cycles in the paths in the conditions and constraint, if the resulting rule is valid and the resulting policy still covers all of $\spa_0$.  A cycle is a path that navigates from some class $C$ back to class $C$.  For example, for the class model in Figure \ref{fig:emr}, the path ``physician.consultations.patient'' contains the cycle ``physician.consultations'', which navigates from Consultation back to Consultation.

% separating (3) from (4) helps us get 100% on the project-management policy. I did not test with e-doc policies, so I dont know if it makes any difference there. it has no effect on the other policies. 

% We have inheritance relationships in class diagram, so we need to generalize rules with objects are subclasses of a same superclass. Our approach: at least 2 of the subclasses have the rule, and there is no invalid UP on other subclasses after generalization.

  The function \hypertarget{mergerulesinher}{\mergerulesinher(\Rules)} attempts to merge a set of rules if their subject types or resource types have a common superclass and all the other components of the rule are the same.  In this case, it replaces that set of rules with a single rule whose subject type or resource type is the most general superclass for which the resulting rule is valid, if any.  For example, rules $\tuple{st_1, sc, rt, rc, c, A}$ and $\tuple{st_2, sc, rt, rc, c, A}$ are replaced with $\rhomerge=\tuple{st', sc, rt, rc, c, A}$ if $\rhomerge$ is valid, and $st'$ is a superclass of $st_1$ and $st_2$, and these conditions do not hold for any superclass of $st'$.

% For example, if we have two rules "A technician can view an one-time work order that is assigned to him", "A technician can view a recurrent work order that is assigned to him", and OnetimeWorkOrder and RecurrentWorkOrder extend superclass WorkOrder in our class model, then we will merge these two rules into rule: "A technician can view a work order that is assigned to him". 

\myparagraph{Optimization}

Our implementation includes two optimizations not reflected in the pseudocode.   (1) Meanings of atomic conditions, conjunctions of atomic conditions, atomic constraints, conjunctions of atomic constraints, and rules are cached are re-used.  (2) The first loop in Figure \ref{fig:alg} processes seed tuples in batches of 1000, and calls $\mergerulesL$ on the rules generated from each batch of seed tuples before adding the resulting rules to $\Rules$.  This reduces the size of $\Rules$ at the end of that loop.  This reduces the overall running time, because $\mergerulesL$ is quadratic in the number of rules.

\subsection{Example}
\label{sec:algorithm:example}

\begin{figure}[htb]
  \centering
\articleonly{\includegraphics[width=0.7\textwidth]{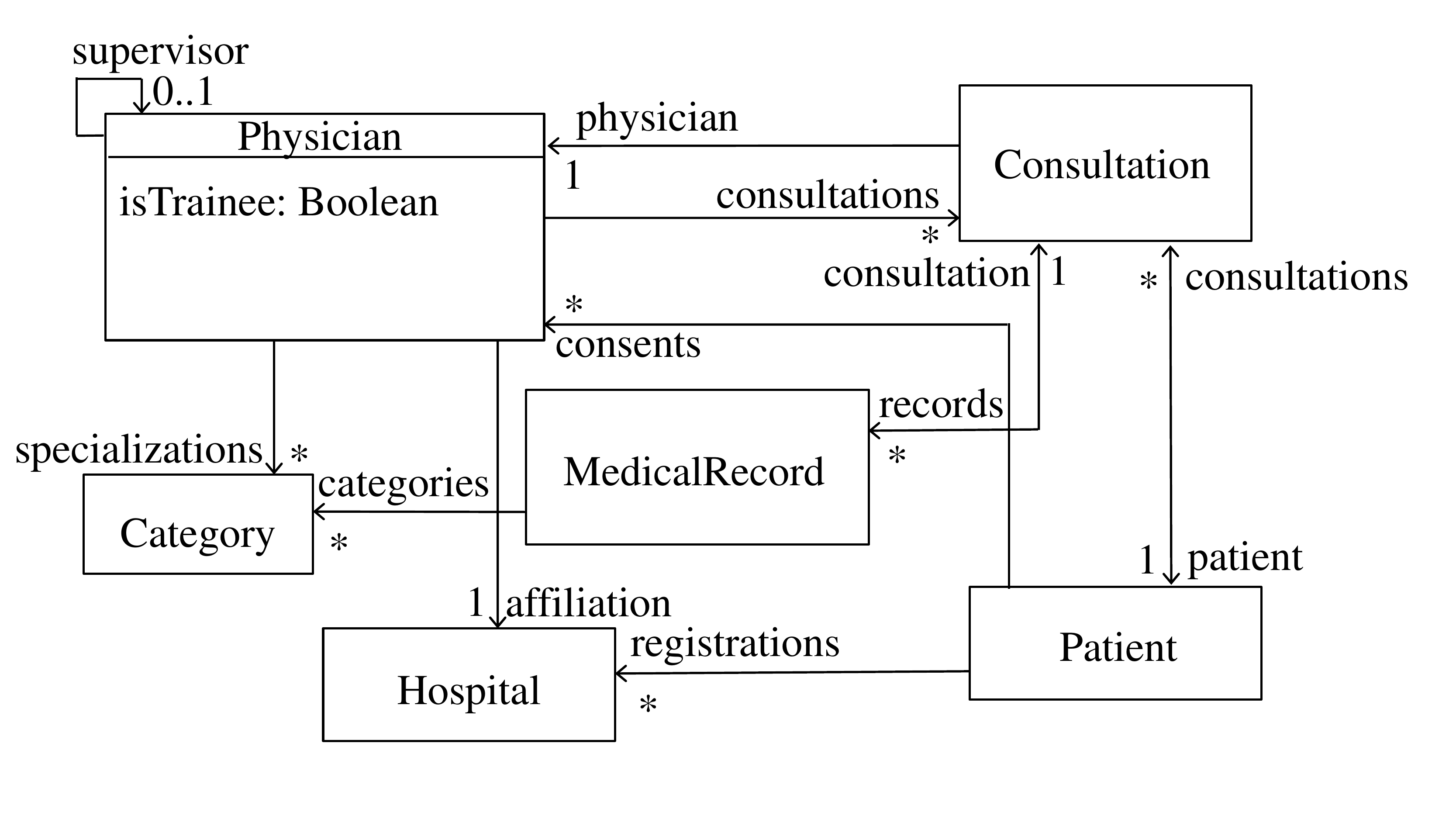}}
  \svonly{\includegraphics[width=0.9\textwidth]{electronic-medical-records}}
  \acmonly{\includegraphics[width=0.9\textwidth]{electronic-medical-records}}
  \lncsonly{\includegraphics[width=0.9\textwidth]{electronic-medical-records}}
  \caption{Class model for Electronic Medical Records (EMR) policy.}
  \label{fig:emr}
\end{figure}

We illustrate the algorithm on a fragment our {\em Electronic Medical Record (EMR) sample policy}, a ReBAC policy based on the EBAC policy in \cite{bogaerts15entity}.  It controls access by physicians and patients to electronic medical records, based on institutional affiliations, patient-physician consultations (each EMR is associated with a consultation), supervisor relationships among physicians, etc.  The class model is in Figure \ref{fig:emr}.  We developed a pseudorandom algorithm that creates object models of varying size for this policy; the algorithm has a size parameter $N$, and the numbers of physicians, patients, consultations, medical records, and hospitals are proportional to $N$.  We generated a object model with $N=15$ for this example.  When describing the execution of the algorithm, we refer to objects by the value of the ``id'' field.  We use id's such as phy0, phy1, $\ldots$ for instances of Physician; consult0, consult1, $\ldots$ for instances of Consultation; and so on.

% $N$ is the number of physicians

% these id's are the same as in the actual object model, except that it uses doc<i> instead of phy<i>.

The policy contains 6 rules.  To keep this example small, we consider here only one rule, namely, the rule in Equation \ref{eq:emr-rule}.

% SUBJECT: subject isInstance Physician and subject.isTrainee = false
% RESOURCE: resource isInstance Consultation
% CONSTRAINT: subject.affiliation.id in resource.patient.registrations.id
%        and subject.id = resource.physician.id
% ACTIONS: createMedicalRecord

Our algorithm selects subject-permission tuple $\langle {\rm phy0}, {\rm consult8},$ ${\rm createMedicalRecord} \rangle$ as the first seed, and then calls $\candidateconstraintL$ to compute the set $\cc$ of atomic constraints that hold between {\rm phy0} and {\rm consult8}.  $\cc$ includes
$c_1$, $c_2$, and $c_3$ where
\begin{eqnarray*}
  c_1 &=& \mbox{subject = resource.physician}\\
  c_2 &=& \mbox{subject.affiliation $\in$ resource.patient.registrations}\\
  c_3 &=& \mbox{subject.affiliation $\in$ resource.patient.consents.affiliation}.
\end{eqnarray*}
The first call to $\addcandidateruleL$ calls $\computecond$ to compute a condition $sc$ that characterizes the set of subjects with permission $\langle {\rm createMedicalRecord},$ ${\rm consult8}\rangle$ and the same candidate constraint as {\rm phy0} for {\rm consult8}.  Condition $sc$ contains the conjunct subject.isTrainee = false along with conjuncts, such as subject.consultations.physician.id = phy0, later removed by simplification.  The second call to $\computecond$ returns a condition $rc$ that characterizes $\set{{\rm consult8}}$; it contains conjuncts such as resource.physician.id = phy0, which is later removed by $\generalizeruleL$, and resource.physician.affiliation.id = hosp1, which is later removed by simplification.

$\addcandidateruleL$ creates rule $\rho_1=\langle {\rm Physician}, sc, {\rm MedicalRecord}$, $rc, \emptyset,$ $\set{\rm createMedicalRecord} \rangle$ and then calls $\generalizeruleL$, which sorts $\cc$ and then attempts to add the atomic constraints in it to $\rho_1$, removing conjuncts for some of the paths used in them.  In one of the recursive calls, the current rule $\rho_2$ contains $c_1$ and $c_2$.  Adding $c_3$ (which uses the same subject path as $c_2$) to this rule worsens rule quality, so $\generalizeruleL$ returns $\rho_2$.  To see the importance of sorting $\cc$, note that, if the algorithm had added $c_1$ and $c_3$ before trying to add $c_2$, then $\generalizeruleL$ would return a rule containing $c_1$ and $c_3$, worsening overall policy quality.  The second call to $\addcandidateruleL$ generates the same candidate rule as the first call, because {\rm createMedicalRecord} is the only action phy0 can perform on consult8.  The algorithm generates other candidate rules from other seeds, and then calls $\mergerulesL$, which merges $\rho_1$ with rules created from permissions of other physicians to create medical records associated with other consultations.  The merged rule is simplified by $\simplifyrulesL$ to produce the desired rule $\rho$.

\section{Evolutionary Algorithm}
\label{sec:evolutionary-alg}

\newcommand{\mergeandsimplifyEvol}{{\rm mergeRulesAndSimplify2}}

Our grammatical evolution algorithm uses the Context-Free Grammar Genetic Programming (CFGGP) approach, in which individuals (which in our context are ReBAC rules) are represented directly as derivation trees of a context-free grammar (CFG).  This is simpler than alternative approaches in which derivation trees are encoded as, e.g., binary strings.  

Classical CFGGP uses two genetic operators to evolve derivation trees: (1) a mutation operator that randomly selects a non-terminal in the derivation tree being evolved, and replaces the existing subtree rooted at that non-terminal with a new subtree randomly generated starting from that non-terminal, and (2) a cross-over operator that randomly selects a non-terminal that appears in both of the derivation trees being evolved (called ``parents''), and swaps the subtrees rooted at that non-terminal.

Our algorithm uses these two classical CFGGP genetic operators (with slight variations, to reflect the focus of the evolutionary search on rules that cover a given seed tuple).  However, we found that the algorithm, with these genetic operators alone, gave poor results, because some mutations that are especially useful in our setting had very low probability.  We solved this problem by introducing additional mutation operators.  For example, we introduced a {\em double mutation} operator, that mutates two out of the three predicates (the subject condition, resource condition, and constraint) in a rule.  This enables the operator to have an effect similar to \generalizeruleL, which changes at least one condition and the constraint.  The same effect can be achieved by two separate mutations to the same rule, but the probability of achieving it that way is much lower.  Although these new mutation operators are specialized to ReBAC policy mining, they are independent of details of our policy language and hence should be equally applicable and effective to extensions of the policy language with additional primitive datatypes (numbers, sequences, etc.) and relational operators (numeric inequality, prefix-of, etc.).

As sketched in Section \ref{sec:intro}, our evolutionary algorithm has two phases: phase 1 constructs a candidate policy, and phase 2 tries to improve the candidate rules by further mutating them.  The improvement phase might seem redundant, because it uses essentially the same mutations as the first phase.  The key difference is that, in phase 1, the benefit of a mutation is evaluated by its effect on rule quality, and in phase 2, it is evaluated in the context of the entire candidate policy by its effect on policy quality.  For example, consider a mutation that transforms a candidate rule $\rho$ into $\rho'$, such that $\rho'$ covers fewer subject-permission tuples, has lower WSC, and has lower rule quality.  If this mutation occurs in phase 1, $\rho'$ might survive, but it is likely to be discarded, due to its lower rule quality.  If this mutation occurs in phase 2, and if the tuples covered by $\rho$ and not by $\rho'$ are also covered by other rules in the candidate policy, $\rho'$ will definitely replace $\rho$ in the candidate policy, because this change reduces the policy's WSC and does not change the policy's meaning.

%TB: The offspring rule has smaller size and cover less tuples in the subject-permission relation. Replacing parent rule with the offspring rule can still improve the policy quality (decreasing policy's WSC) if all tuples covered by the parent rule and not by the offspring rule are covered by other rules in the policy. In other words,  we can perform the replacing if the meaning of the policy does not change. If we could generate the same offspring in the search step, this rule may be discarded since its fitness value is less than the parent's.

%\paragraph{Grammar generation.}

{\it Grammar generation} is performed before the main part of the evolutionary algorithm, to specialize the generic grammar of ORAL to a specific input.  The language of the generated grammar contains rules satisfying the restrictions: (1) constants are limited to those appearing in the object model, (2) class names and field names are limited to those appearing in the class model, (3) paths in conditions and constraints are type-correct, based on the class model, and satisfy the same length limits as in the greedy algorithm, and (4) actions are limited to those appearing in the given subject-permission relation.  The grammar generation algorithm pre-computes all atomic conditions and atomic constraints satisfying these restrictions.  For a type $t$, let $\set{c_{t,1}, c_{t,2}, \ldots, c_{t,n_t}}$ denote the set of atomic conditions on objects of type $t$ that satisfy these restrictions.

The starting non-terminal $N_{\it rule}$ has alternatives corresponding to rules with different subject and resource types.  Specifically, each alternative for $N_{\it rule}$ is a rule with two non-terminals: a non-terminal $N_{\it rule(t_1,t_2)}$ that generates all components of a rule with subject type $t_1$ and resource type $t_2$ except for the action set component (which is independent of the types), and a non-terminal $N_{\it act}$ that generates subsets of the actions that appear in $\spa_0$.  Each alternative for $N_{\it rule(t_1,t_2)}$ contains non-terminals $N_{\it cond(t_1)}$ and $N_{\it cond(t_2)}$ to generate the subject condition and resource condition, respectively, and a non-terminal $N_{\it cons(t_1,t_2)}$ to generate the constraint.  The non-terminal $N_{\it cond(t)}$ generates conditions on objects of type $t$.  Specifically, it generates a sequence of non-terminals $N_{t,1}, N_{t,2}, \ldots, N_{t,n_t}$, where each $N_{t,i}$ can generate either atomic condition $c_{t,i}$ or the empty string; this allows the condition to contain an arbitrary subset of the atomic conditions on objects of type $t$.  The non-terminal $N_{\it cons(t_1,t_2)}$ generates constraints relating objects of types $t_1$ and $t_2$; the productions for it are defined in a similar way as the productions for conditions.

% old version in which N_act is a child of N_{rule(t1,t2)}
%The production for the starting non-terminal $N_{\it rule}$ for rules has alternatives of the form $N_{\it rule(t_1,t_2)}$, where $t_1$ and $t_2$ are types (classes), and the non-terminal $N_{\it rule(t_1,t_2)}$ generates rules with subject type $t_1$ and resource type $t_2$.  The production for $N_{\it rule(t_1,t_2)}$ expands it to a rule containing non-terminals $N_{\it cond(t_1)}$ and $N_{\it cond(t_2)}$ for the subject condition and resource condition, respectively, non-terminal $N_{\it cons(t_1,t_2)}$ for the constraint, and non-terminal $N_{\it act}$ for the set of actions.  The non-terminal $N_{\it cond(t)}$ generates conditions on objects of type $t$.  Specifically, it expands to a sequence of non-terminals $N_{t,1}, N_{t,2}, \ldots, N_{t,n_t}$, where each $N_{t,i}$ can expand to either atomic condition $c_{t,i}$ or the empty string; this allows the condition to contain an arbitrary subset of the atomic conditions on objects of type $t$  The non-terminal $N_{\it cons(t_1,t_2)}$ generates constraints relating objects of types $t_1$ and $t_2$; the productions for it are defined in a similar way as the productions for conditions.  The production for $N_{\it act}$ has alternatives corresponding to subsets of the actions that appear in $\spa_0$.

% if we restore the following, need to update it to take inheritance into account.
% subsets of $\setc{a}{\tuple{s,r,a}\in\spa_0 \,\land\, \type(s)=t_1 \,\land\, \type(r)=t_2}$ (the set of actions that some subject of type $t_1$ can perform on some resource of type $t_2$).

\begin{figure}[tbp]
\begin{tabular}[t]{@{}l@{}}
// \com{Phase 1: Construct candidate policy, using evolutionary search to find one rule at a time.}\\
$\Rules = \emptyset$\\
$\uncovsp = \spa_0.\Copy()$\\
\whileloop\ $\neg\uncovsp.\isempty()$\\
\ind $\tuple{s,r,a}$ = highest-quality tuple in $\uncovsp$ using $\QspL$ metric ~~~ // \com{seed for this iteration} \\
\ind {\it pop} = initialPopulation($\tuple{s,r,a}, \Rules, \uncovsp$)\\
\ind \forloop\ ${\it gen}$ = 1 to {\it nGenerationsSearch}\\
\ind\ind $\op$ = an operator selected from {\it searchOps} using probability distribution {\it searchOpDist}\\
\ind\ind $S$ = set of {\it nTournament} rules randomly selected from {\it pop}\\
\ind\ind \ifstmt\ $op$ is a mutation\\
\ind\ind\ind {\it pop}.add(the rule generated by applying $op$ to the highest-quality rule in $S$)\\
\ind\ind\elsestmt\ ~~~ // \com{$op$ is a cross-over}\\
\ind\ind\ind {\it pop}.add(the two rules generated by applying $op$ to the two highest-quality rules in $S$)\\
\ind\ind\eifstmt\\
\ind\ind remove the lowest-quality rules in {\it pop} until $\vert {\it pop}\vert$ = popSize\\
\ind \eforloop\\
\ind $\rho$ = the highest-quality rule in {\it pop}\\
\ind \ifstmt\ $\valid{\rho}$\\
\ind\ind \Rules.add($\rho$)\\
\ind\ind \uncovsp.removeAll($\mean{\rho}$)\\
\ind \eifstmt\\
\ewhileloop\\
// \com{Phase 2: Improve the candidate rules by further mutating them.}\\
\foreachloop\ $\rho$ \forloopin\ \Rules\\
\ind\forloop\ ${\it gen}$ = 1 to {\it nGenerationsImprove}\\
%\ind\ind // \com{If no improvements are found in the first {\it nGenerationsImprove}/2 attempts, give up.}\\
\ind\ind \ifstmt\ {\it gen} = {\it nGenerationsImprove}/2 $\land$ (all attempted improvements to $\rho$ failed)\\
\ind\ind\ind \breakstmt\\
\ind\ind\eifstmt\\
\ind\ind $\op$ = an operator selected from {\it improveOps} using probability distribution {\it improveOpDist}\\
\ind\ind $\rho'$ = the rule generated by applying $op$ to $\rho$\\
\ind\ind \ifstmt\ ${\rm wellFormed}(\rho') \,\land\, \valid{\rho'} \,\land\, \ID(\rho')\le\ID(\rho)$\\
\ind\ind\ind ${\it redundant} = \setc{\rho_0\in\Rules}{\mean{\rho_0}\subseteq\mean{\rho'}}$\\
\ind\ind\ind\ifstmt\ $(\Rules\union\set{\rho'}\setminus {\it redundant})$ covers $\spa_0$ and has lower WSC than $\Rules$\\
\ind\ind\ind\ind \Rules.removeAll({\it redundant})\\
\ind\ind\ind\ind \Rules.add($\rho'$)\\
\ind\ind\ind\eifstmt\\
\ind\ind\eifstmt\\	
\ind\eforloop\\
\eforloop\\
%Note: we use rule quality metric from ReBAC Miner for mergeAndSimplify
\mergeandsimplifyEvol(\Rules)\\
\return\ \Rules
\end{tabular}
\caption{Evolutionary algorithm for ReBAC policy mining.  Inputs: subject-permission relation $\spa_0$, class model $\cm$, and object model $\om$.  Output: set of rules $\Rules$.  The algorithm also has numerical paramters that determine population size, number of generations, etc., as described in the text.}
\label{fig:evol-alg}
\end{figure}

Pseudocode for the main part of our evolutionary algorithm appears in Figure \ref{fig:evol-alg}.  All random choices follow a uniform distribution, unless a different probability distribution is specified.  In our experiments, values of the numerical parameters are: {\it popSize}=200, {\it nGenerationsSearch}=2000, {\it nTournament}=15, {\it nGenerationsImprove} = 1000.  These values, and other numerical parameter values mentioned below, were selected based on tuning experiments, described in Section \ref{sec:evaluation}.

% from Eric: The parameter values popSize=200 and nTournament=15 essentially correspond to a very high selection pressure which, itself, greatly favors the exploitation in spite of the exploration. I agree that it is unfeasible to study in depth the impact of the many evolutionary parameters and I have not elements for saying that a different choice could lead to better results. However, a "evolutionary reviewer" might raise this issue; my answer would be something like this: we chose to favor exploitation because we built the initial population such that it includes reasonably good solutions.
% another reason for favoring selection over exploration is that we don't just use random initialization of the population, we use relatively smart initialization that should already get us close to the desired rules.
 
% from Eric: (Concerning the tournament size: values larger than 10% of the population are quite unusual).

Function initialPopulation($\tuple{s,r,a}, \Rules, \uncovsp$) creates an initial population for the evolutionary search for a high-quality rule that covers the seed $\tuple{s,r,a}$ and other tuples.  It is implicitly parameterized by the desired population size {\it popSize}. Half of the desired rules are generated using method 1; the other half are generated using method 2.

Method 1 (candidate rules generated as in greedy algorithm plus random variants): Generate candidate rules from {\it seed}, $\uncovsp$, and $\Rules$ in the same way as the two calls to $\addcandidateruleL$ in Figure \ref{fig:alg} and add them to the initial population; this ensures that it contains at least 1 valid rule that covers at least one uncovered tuple.  To generate the remaining rules, repeatedly randomly select a rule currently in the initial population, remove randomly selected atomic conditions from the subject condition until the number of remaining atomic conditions equals a target number randomly selected in the interval 1..7, do the same for the resource condition, remove randomly selected atomic constraints until the number of remaining atomic constraints equals a target number randomly selected in the interval 1..3, and add the resulting rule to the initial population.

Method 2 (random candidate rules): Each rule has subject type $\type(s)$ or one of its ancestors, resource type $\type(r)$ or one of its ancestors, subject condition and resource condition selected as described below, randomly selected constraint consistent with the selected subject type and resource type, and action set $\set{a}$.  If $\type(s)$ has an ancestor, then the probability of using $\type(s)$ as the subject type is 0.8, and the probability of using an ancestor (selected uniformly at random among the ancestors) is 0.2; similarly for the resource type.  To generate the subject condition and resource condition, we randomly select among the following three cases, and then select randomly select a condition within the selected case: no condition (i.e., the condition is always true), a condition on a single-valued path, and an arbitrary condition.

% we simplify the fitness function slightly by omitting the denominators from FAR and FRR.  they are rule-independent hence don't affect results of comparisons.

Rule quality is measured using the same fitness function $f$ as \cite{medvet2015} (our definition is slightly simplified but equivalent):
$f(\rho) = \tuple{\far(\rho),\frr(\rho),\ID(\rho),\wsc(\rho)}$, where the {\em false acceptance rate} is $\far(\rho)=|\mean{\rho}\setminus\uncovsp|$, the {\em false rejection rate} is $\frr(\rho)=|\uncovsp\setminus\mean{\rho}|$, and $\ID(\rho)$ equals 2 if the subject condition and resource condition both contain an atomic condition with path ``id'', equals 1 if exactly one of them does, and equals 0 if neither of them does.  The fitness ordering is lexicographic order on these tuples, where smaller is better.

The set of genetic operators used in the search phase, denoted {\it searchOps}, contains: (1) single mutation: first, randomly select whether to mutate the subject condition, resource condition, or constraint, then randomly select a non-terminal $N$ in that part of the derivation tree, and then randomly re-generate the subtree rooted at $N$; (2) double mutation: same as single mutation, except, in the first step, choose two out of the three possibilities, and then perform the remaining steps for both of them; (3) action mutation: in the action set component of the rule, randomly add or remove actions that subject $s$ can perform on $r$ according to $\spa_0$, subject to the restriction that we never remove action $a$, where $\tuple{s,r,a}$ is the seed tuple for this search; (4) simplify mutation: remove one randomly selected atomic condition (from the subject condition or resource condition) or atomic constraint; (5) crossover: randomly select a non-terminal $N$ in the subtree for the subject condition, resource condition, or constraint in one parent, find the same non-terminal in the other parent (if it does not appear, select a different non-terminal in the first parent), and swap the subtrees rooted at those two occurrences of $N$.

We describe the genetic operators as if they directly manipulate abstract syntax trees, because this allows a higher-level and more intuitive presentation.  However, the genetic operators actually manipulate derivation trees of the generated grammar.

% searchOpDist: 0.1 for crossover, approx 0.24 each for single, action, and simplify mutation, approx 0.17 for double mutation.  these sum to 0.99.

{\it searchOpDist} specifies the probability of selecting each genetic operator in {\it searchOps}.  First, it selects the type of genetic operator, selecting mutation with probability 0.9, or crossover with probability 0.1.  If mutation is selected, the probability of selecting each of the four types of mutation is proportional to its weight, where single mutation, action mutation, and simplify mutation each have weight 1, and double mutation has weight 0.7.

The set of genetic operators used in the improvement phase, denoted {\it improveOps}, contains: (1) single mutation; (2) double mutation; (3) type+single mutation: randomly select whether to replace the subject type, resource type, or both with their parent types (if those parents exist), apply a single mutation, check whether the resulting rule is well-formed (because the unchanged condition or constraint might be inconsistent with the changed type), and if not, discard it; (4) type+double mutation: same as type+single mutation, except with a double mutation instead of a single mutation.

{\it improveOpDist} specifies the probability of selecting each genetic operator in {\it improveOps}. The probabilities are: single mutation, 0.09; double mutation, 0.81; type+single mutation, 0.01; type+double mutation, 0.09.

Function $\mergeandsimplifyEvol$ is the same as $\mergeandsimplifyL$ except that it incorporates one additional simplification: replace the subject type or resource type with a child of that type, if the policy still covers $\spa_0$.

% the need for this simplification (i.e., a rule whose type is ``too general'') did not arise in experiments with greedy alg, so we did not bother to introduce it there.
% similarly, we don't call mergeRulesInheritance in evolutionary alg because the need for it (i.e., candidate policy containing two rules with sibling types that need to be merged into one rule with parent type) did not arise in experiments with evolutionary alg.

% !TeX root = main.tex

\section{Sample Policies and Case Studies}
\label{sec:sample-policies}

% the policies are in src\RelationalPolicyMining\experiment results\full-experiment-02-19\ in the "data" folder and echoed in the output file.

% the policies are described in txt files in policy-mining, except grant proposal, described in
% policy-mining\grammatical evolution\sample-policies\rebac\grant-proposal\grant-proposal.txt

We developed four sample policies, which have non-trivial and realistic rules, but are relatively small.  We also translated two large case studies into ORAL.  They were developed by Decat, Bogaerts, Lagaisse, and Joosen based on the access control requirements for Software-as-a-Service (SaaS) applications offered by real companies \cite{decat14edoc,decat14workforce}.  We translate their detailed natural-language descriptions of the policies into class models and ReBAC rules, omitting a few aspects left for future work, mainly temporal conditions, obligations, and policy administration. 
% The third case study is for a university's grant proposal workflow system \cite{grantproposal}.

Each policy has handwritten class model and rules, and a synthetic object model generated by a policy-specific pseudorandom algorithm designed to produce realistic object models, by creating objects and selecting their attribute values using appropriate probability distributions (e.g., normal, uniform, and Zipf distributions).  The object model generation algorithm for each policy is parameterized by a size parameter $N$; for most classes, the number of instances is selected from a normal distribution whose mean is linear in $N$.  Figure \ref{fig:policy-size} shows several metrics of the size of the rules, class model, and object model in each policy. 

% our small sample policies have fewer conditions and more constraints, compared to the larger case studies.

\begin{table*}[htb]
  \centering
\begin{tabular}{|l|l|l|l|l|l|l|l|l|l|l|l|l|}
\hline
Policy & \#rules & \#cond/rule & \#constr/rule & \#classes & $N$ & \multicolumn{1}{c|}{\#obj} & \multicolumn{1}{c|}{\#field/obj} & \multicolumn{1}{c|}{$|\spa_0|$}\\ 
\hline
EMR & 6 & 0.17 & 1.3 & 6 & 15 & 344 & 3.5 & 708\\ 
\hline
healthcare & 9 & 0 & 1.1 & 12 & 5 & 737 & 3.5 & 2207\\ 
\hline
project mgmt. & 13 & 0.08 & 1.2 & 15 & 5 & 181 & 2.7 & 322\\ 
\hline
university & 10 & 0.40 & 0.70 & 10 & 5 & 731 & 2.2 & 2439\\ 
\hline
e-document & 39 & 2.3 & 0.59 & 16 & 125 & 421 & 5.9 & 2687\\ 
\hline
workforce mgmt. & 27 & 1.7 & 0.63  & 29 & 10 & 411 & 3.7 & 1739\\ 
\hline
\end{tabular}
\caption{Policy sizes. \#cond/rule and \#constr/rule are the average numbers of conditions per rule and constraints per rule, respectively.  For the given value of $N$, \#obj is the average number of objects in the object model, and \#field/obj is the average number of fields (including ``id'' field) per object in the object model. Averages are over 30 pseudorandom object models for each policy.}
  \label{fig:policy-size}
\end{table*}

% full values for rule complexity metrics:
%EMR_15:
%Average number conditions per rule: 0.16666666666666666
%Average number constraints per rule: 1.3333333333333333
%e-doc_125:
%Average number conditions per rule: 2.2564102564102564
%Average number constraints per rule: 0.5897435897435898
%eWorkforce_10:
%Average number conditions per rule: 1.7037037037037037
%Average number constraints per rule: 0.6296296296296297
%grantProposal_50:
%Average number conditions per rule: 1.5769230769230769
%Average number constraints per rule: 0.6153846153846154
%healthcare_5:
%Average number conditions per rule: 0.0
%Average number constraints per rule: 1.1111111111111112
%project-management_5:
%Average number conditions per rule: 0.08333333333333333
%Average number constraints per rule: 1.1666666666666667
%university_5:
%Average number conditions per rule: 0.4
%Average number constraints per rule: 0.7

% average total number of fields in all objects in an object model, excluding id field
%EMR 854
%healthcare  1806
%project mgmt. 300
%university  908 
%e-document  2045
%workforce mgmt.  1123

% choose sample rules with long paths

The {\em Electronic Medical Record (EMR) sample policy} is described in Section \ref{sec:algorithm:example}.
% Another sample rule is ``A physician at a facility can view a medical record for a consultation with any physician at that facility by a patient still registered at the facility'', expressed as $\langle\,$Physician, true, MedicalRecord, true, subject.affiliation = resource.consultation.physician.affiliation $\land$ subject.affiliation $\in$ resource.consultation.patient.registrations, \{view\}$\rangle$.

The {\em healthcare sample policy}, based on the ABAC policy in \cite{xu15miningABAC}, controls access by nurses, doctors, patients, and agents (e.g., a patient's spouse) to electronic health records (HRs) and HR items (i.e., entries in health records).
%, based on clinicians' membership in medical teams and wards, the teams treating each patient, physicians' medical specialties, patient-agent relationships, the authorship and topic of each HR item, etc.  
The numbers of wards, teams, doctors, nurses, teams, patients, and agents are proportional to $N$.

% $N$ is the number of wards

The {\em project management sample policy}, based on the ABAC policy in \cite{xu15miningABAC}, controls access by department managers, project leaders, employees, contractors, auditors, accountants, and planners to budgets, schedules, and tasks associated with projects.
%, based on the department each project is in, the manager of each department, the leader of each project, the tasks assigned to each worker, the areas of expertise required for each task, whether each task is proprietary, etc.  
The numbers of departments, projects, tasks, and users of each type are proportional to $N$.  

%A sample rule appears in Section \ref{sec:language}.  
%Another sample rule is ``The manager of a department can read and approve the budget for a project in the department'', expressed as $\langle\,$ Manager, true, Budget, true, subject.department = resource.project.department, \{read, approve\}$\rangle$, where Budget.project is the project that the budget is for.

% $N$ is the number of departments

The {\em university sample policy}, based on the ABAC policy in \cite{xu15miningABAC}, controls access by students, instructors, teaching assistants (TAs), department chairs, and staff in the registrar's office and admissions office to applications (for admission), gradebooks, transcripts, and course schedules.
%, based on the courses each faculty teaches, the courses each student is TA for, each student's major, the courses each student takes, the chair of each department, the office each staff works in, etc.  
The numbers of departments, students, faculty, and applicants for admission are proportional to $N$.
%\fullonly{  A sample rule is ``The chair of a department can read the transcripts of all students in that department'', expressed as $\langle\,$Faculty, subject.isChair = true, Transcript, true, subject.department = resource.student.major, \{read\}$\rangle$, where Transcript.student is the student that the transcript is for.}

Rewriting the preceding three policies in ReBAC allows numerous aspects to be expressed more naturally than in ABAC.  This is reflected in rules that use paths with length greater than one, not counting occurrences of ``id''.  For example, consider the constraint ``subject.teams contains resource.record.patient.treatingTeam'' in the above example rule from the healthcare policy.  In the ReBAC policy, ``treatingTeam'' is, naturally, an attribute of Patient.  In the original ABAC policy, there is no way to navigate from the HR item to the patient; to circumvent this limitation, a patient's ``treatingTeam'' attribute is (unnaturally) duplicated in each HR item in each HR for that patient.  
%As another example, consider the constraint ``subject.department = resource.project.department'' in the above example rule for project management.  In the ABAC policy, there is no way to navigate from a budget to the associated project, so each project's ``department'' attribute is duplicated in the project's budget.

% $N$ is the number of departments

The {\em e-document case study}, based on \cite{decat14edoc}, is for a SaaS multi-tenant e-document processing application.  The application allows tenants to distribute documents to their customers, either digitally or physically (by printing and mailing them). 
% There are rules for various types of employees of the e-document company.  
The overall policy contains rules governing document access and administrative operations by employees of the e-document company, such as helpdesk operators and application administrators.  It also contains specific policies for some sample tenants.
One sample tenant is a large bank, which controls permissions to send and read documents based on (1) employee attributes such as department and projects, (2) document attributes such as document type,
%(e.g., invoice, sales offer, banking note
related project (if any), and presence of confidential or personal information, and (3) the bank customer
% (which may be an individual or an organization) 
to which the document is being sent.  
Some tenants have semi-autonomous sub-organizations, modeled as sub-tenants, each with its own specialized policy rules.  
%For example, the large bank's leasing business is a sub-tenant.  Another sample tenant is a news agency, which controls permissions to send and read documents based on employee attributes such as department, position, and regional office, and document attributes such as document type (e.g., contract, paycheck, invoice).
The numbers of employees of each tenant, registered users of each customer organization, and documents are proportional to $N$.  

% another sample rule with a long path:
% - Members of the helpdesk can only read the content of documents belonging to tenants for which they are assigned responsible.
% SUB: sub isInstance HelpdeskOperator
% RES: res isInstance Document and res.isConfidential = false
% CON: sub.tenants.id contains res.sentBy.employer.id
% ACT: read

% N is the number of employees of all tenants collectively.

The {\em workforce management case study}, based on \cite{decat14workforce}, is for a SaaS workforce management application provided by a company, pseudonymously called eWorkforce, that handles the workflow planning and supply management for product or service appointments (e.g., install or repair jobs).  Tenants (i.e., eWorkforce customers) can create tasks on behalf of their customers.
%; typically this is done by helpdesk operators working either for the tenant itself or for a helpdesk supplier to which the tenant outsourced its helpdesk.
Technicians working for eWorkforce, its workforce suppliers, or subcontractors of its workforce suppliers receive work orders to work on those tasks, and appointments are scheduled if appropriate.  Warehouse operators receive requests for required supplies. 
% The complexity of the class model in Figure \ref{fig:workforce} reflects the complexity of the policy. 
The overall policy contains rules governing the employees of eWorkforce, as well as specific policies for some sample tenants, including PowerProtection (a provider of power protection equipment and installation and maintenance services) and TelCo (a telecommunications provider, including installation and repair services).  Permissions to view, assign, and complete tasks are based on each subject's position,
% (manager, application administrator, technician, sales manager, helpdesk operator, warehouse operator, etc.), 
the assignment of tasks to technicians, the set of technicians each manager supervises, the contract (between eWorkforce and a tenant) that each work order is associated with, the assignment of contracts to departments within eWorkforce, etc.  The numbers of helpdesk suppliers, workforce providers, subcontractors, helpdesk operators, contracts, work orders, etc., are proportional to $N$.

%\begin{figure}[tb]
%  \centering
%  \includegraphics[width=\textwidth]{workforce-management}
%  \caption{Class model for workforce management case study.}
%  \label{fig:workforce}
%\end{figure}

% VALUES USED IN IMPLEMENTATION
%For all policies, $\mcse = 5$.  For EMR, $\mspl=3$, $\mrpl=4$, $\sped=0$, $\rped=1$, and $\mtpl=6$.  For healthcare, project management, and university, $\mspl=3$, $\mrpl=3$, $\sped=0$, $\rped=0$, and $\mtpl=6$.  For e-document, $\mspl=4$, $\mrpl=4$, $\sped=0$, $\rped=0$, and $\mtpl=6$.  For workforce management, $\mspl=3$, $\mrpl=3$, $\sped=0$, $\rped=2$, and $\mtpl=7$.

% these path lengths include trailing "id" field.  for conditions, we should  count id, so \mspl and \mrpl are fine.  for constraints, we should not  count id.  currently, we mine only non-boolean constraints, so in every constraint, the subject path and resource path both end with id, so if we don't count id, we can reduce \mtpl by 2.  \sped and \rped are unaffected, because they are relative amounts ("extra distance"), not absolute path lengths.

The algorithm parameters are set as follows in our experiments.  For all policies, $\mcse = 5$.  For EMR, $\mspl=3$, $\mrpl=4$, $\sped=0$, $\rped=1$, and $\mtpl=4$.  For healthcare, project management, and university, $\mspl=3$, $\mrpl=3$, $\sped=0$, $\rped=0$, and $\mtpl=4$.  For e-document, $\mspl=4$, $\mrpl=4$, $\sped=0$, $\rped=0$, and $\mtpl=4$.  For workforce management, $\mspl=3$, $\mrpl=3$, $\sped=0$, $\rped=2$, and $\mtpl=5$.  The parameter values are similar across policies, though they vary by 1 or 2.  A reasonable parameter value selection strategy is to start with values similar to these, perhaps on the lower end, and increase them slightly if the mined policy is unsatisfactory.

\section{Evaluation}
\label{sec:evaluation}

To evaluate the effectiveness of our algorithms, we start with a ReBAC policy, generate ACLs representing the subject-permission relation, run our algorithms on the ACLs along with the class model and object model, and compare the mined ReBAC policy with the policy produced by applying $\simplifyrulesL$ to the original policy; we refer to the latter as the {\em simplified original policy}.  If the mined policy is similar to the simplified original policy, the policy mining algorithm succeeded in discovering the rules that are implicit in the ACLs.  Comparison with the simplified original policy is a more robust measure of the algorithm's ability to discover high-level rules than comparison with the original policy, because the original policy is not always the simplest.  For our four sample policies, the simplified original policy is identical to the original policy; for the two large case studies, the simplified original policy has lower WSC than the original policy.

%e-doc: orig WSC: 359  simp orig WSC: 250
%workforce-mgtmt.: orig WSC 262  simp orig WsC: 208

Our algorithm is implemented in Java.  Experiments were run using Java 8 on Windows 10 on an Intel i7-6770HQ CPU.  The code and data are available at \url{http://www.cs.stonybrook.edu/~stoller/software/}.

% jiajie : Intel Core i7-6770HQ Processor, 6M Cache, up to 3.50 GHz
% Thang : Intel Core i7-3770 Processor, 8M Cache, up to 3.90 GHz

\subsection{Policy Similarity Metrics}
\label{sec:metrics}

Both of our policy similarity metrics are normalized to range from 0 (completely different) to 1 (identical).

\myparagraph{Syntactic Similarity}

Syntactic similarity measures the fraction of types, atomic conditions, atomic constraints, and actions that rules or policies have in common.  The Jaccard similarity of sets is $J(S_1, S_2) = |S_1\intersect S_2| \,/\, |S_1 \union S_2|$.  The {\em syntactic similarity of rules} $\rho_1=\langle st_1, sc_1, rt_1,$ $rc_1, c_1, A_1\rangle$ and $\rho_2=\tuple{st_2, sc_2, rt_2, rc_2, c_2, A_2}$ is the average of $J(\set{st_1}, \set{st_2})$, $J(sc_1, sc_2)$, $J(\set{rt_1}, \set{rt_2})$, $J(rc_1, rc_2)$, $J(c_1, c_2)$ and $J(A_1, A_2)$.  The {\em syntactic similarity of rule sets} $\Rules_1$ and $\Rules_2$, {\em SynSim}($\Rules_1$, $\Rules_2$), is the average, over rules $\rho$ in $\Rules_1$, of the syntactic similarity between $\rho$ and the most similar rule in $\Rules_2$. 

\myparagraph{Semantic Similarity}

Semantic similarity measures the similarity of the meanings of rules.  The {\em semantic similarity of rules} $\rho_1$ and $\rho_2$ is $J(\mean{\rho_1},\mean{\rho_2})$. We extend this to {\em rule-wise semantic similarity of policies} {\em RSemSim}($\Rules_1$, $\Rules_2$) exactly the same way that syntactic similarity of rules is extended to syntactic similarity of policies.  {\em Note that this metric measures similarity of the meanings of the rules in the policies, not similarity of the overall meanings of the policies.}  This metric is slightly more abstract than syntactic similarity, because it ignores syntactic differences that do not affect the meaning of a rule, such as including a conjunct that is unnecessary because it is implied by another conjunct or a constraint.

% \myparagraph{Fractions of Under-Assignments and Over-Assignments}

% To characterize the semantic differences between an original ABAC policy
% $\pi_0$ and a mined policy $\pi$ in a way that distinguishes
% under-assignments and over-assignments, we compute the fraction of
% over-assignments and the fraction of under-assignments, defined by
% $|\mean{\pi}\setminus\mean{\pi_0}|\,/\,|\mean{\pi}|$ and
% $|\mean{\pi_0}\setminus\mean{\pi}|\,/\,|\mean{\pi}|$, respectively.

% parameter tuning results are in svnstoller\policy-mining\grammatical evolution\evolved-ge-master\experiment_result\tuning-parameters-experiments.xslx

\myparagraph{Parameter Tuning}
%\label{sec:eval:tuningParameters}

Our evolutionary algorithm has several parameters (population size, number of generations, etc.).  The parameters and their values used in our experiments are presented in Section \ref{sec:evolutionary-alg}.  We determined those values through a series of experiments, in which we started with initial guesses at good values of the parameters, varied one parameter, selected the value that gave the best average policy similarity results, and then proceeded to vary the next parameter (in a somewhat arbitrary order, except that we vary parameters used in initialization before parameters used in phase 1 before parameters used in phase 2).  After varying each parameter once, we varied some parameters again, and found little change in the results, so we did not bother with an exhaustive optimization process that would consider all combinations of values of all parameters.

\subsection{Policy Similarity Results}
\label{sec:eval:similarity}

% policy similarity results are from "re-run testing" tab of policy-mining\grammatical evolution\evolved-ge-master\experiment_result\tuning-parameters-experiments.xlsx

\begin{table*}[tb]
\centering
\begin{tabular}{|l|l|l|l|l|l|l|l|l|l|l|l|}
\hline
Policy
& \multicolumn{4}{c|}{Syntactic Similarity}                & \multicolumn{4}{c|}{Rule-wise Semantic Sim.}                 & \multicolumn{3}{c|}{WSC}                                                                 \\ \cline{2-12} 
                        & \multicolumn{2}{c|}{Evol.} & \multicolumn{2}{c|}{Greedy} & \multicolumn{2}{c|}{Evol.} & \multicolumn{2}{c|}{Greedy} & \multicolumn{1}{c|}{SimpOrig} & \multicolumn{1}{c|}{Evol.} & \multicolumn{1}{c|}{Greedy} \\ \cline{2-12} 
                        & $\mu$      & $\sigma$      & $\mu$       & $\sigma$      & $\mu$       & $\sigma$     & $\mu$       & $\sigma$      & $\mu$                         & $\mu$                      & $\mu$                       \\ \hline
EMR                     & {\bf 0.99} & 0.01          & 0.99        & 0.01          & {\bf 1.00}  & 0.00         & 1.00        & 0.00          & 49                            & 49                         & 50                          \\ \hline
healthcare              & {\bf 1.00} & 0.00          & 1.00        & 0.00          & {\bf 1.00}  & 0.00         & 1.00        & 0.00          & 54                            & 54                         & 54                          \\ \hline
project mgmt.           & {\bf 1.00} & 0.00          & 1.00        & 0.00          & {\bf 1.00}  & 0.00         & 1.00        & 0.00          & 76                            & 76                         & 76                          \\ \hline
university              & {\bf 1.00} & 0.00          & 1.00        & 0.00          & {\bf 1.00}  & 0.00         & 1.00        & 0.00          & 54                            & 54                         & 54                          \\ \hline
e-document              & {\bf 0.90} & 0.03          & 0.86        & 0.02          & {\bf 0.86}  & 0.08         & 0.72        & 0.07          & 250                           & 326                        & 416                         \\ \hline
workforce mgmt.         & {\bf 0.92} & 0.02          & 0.81        & 0.02          & {\bf 0.96}  & 0.02         & 0.90        & 0.03          & 208                           & 185                        & 229                         \\ \hline
\end{tabular}
\caption{Policy similarity results.  Evol. and Greedy refer to the rules mined by the  evolutionary algorithm and greedy algorithm, respectively.  SimpOrig refers to the simplified original rules.  When computing policy similarity, the first argument to {\it SynSim} and {\it RSemSim} is the mined rules, and the second argument is the simplified original rules.  $\mu$ is the mean over 30 pseudorandom object models, and $\sigma$ is the standard deviation.  Similarity results for the evolutionary algorithm are emphasized with bold font, since they are as good as or better than the results for the greedy algorithm in all cases. }
\label{fig:similarity}
\end{table*}

Figure \ref{fig:similarity} shows the results of policy similarity measurements.  The policy size parameter $N$ has the values shown in Figure \ref{fig:policy-size}.  We set all weights $w_i$ in the definition of $\wsc$ to 1.  

For the four sample policies, for both of our policy mining algorithms, the mined policy is identical to the simplified original policy, except for one minor syntactic variation in one conjunct of one condition of one rule of the EMR policy (the variant is semantically equivalent to the original conjunct; this is reflected in the perfect rule-wise semantic similarity).  For the two large case studies, the evolutionary algorithm does better than the greedy algorithm: for e-document, the syntactic similarity and rule-wise semantic similarity are 4\% and 14\% higher, respectively (but the running time is longer, as discussed below); for workforce management, they are 11\% and 6\% higher, respectively.
 
%For the {\em healthcare policy}, {\em project management policy}, and {\em university policy}, the simplified original and mined policies from both algorithms are identical.  For the {\em EMR policy}, the mined policy from both algorithms has perfect rule-wise semantic similarity with the simplified original policy (hence also with the original policy) and nearly perfect (0.99) average syntactic similarity with it.
% up-to-date but omitted to save space.
% The latter reflects that the simplified original and mined policies are identical in some runs and have a syntactic difference in one rule in other runs.  In particular, the mined version of the example rule in Section \ref{sec:algorithm:example} sometimes has the constraint ``subject.consultations contains resource'' instead of ``subject = physician.resource''.  These constraints are semantically equivalent, because the associations used in them ({\it cf.} Figure \ref{fig:emr}) are inverses of each other.

% from full-experiment-02-19\EMR_15\EMR_15_0.output [experiments for SACMAT]
% Input rule: 
% rule(Physician; isTrainee in {false}; Consultation; ; affiliation.id in patient.registrations.id and id = physician.id; {createMedicalRecord})
% Similar output rules: 
% rule(Physician; isTrainee in {false}; Consultation; ; consultations.id contains id and affiliation.id in patient.registrations.id; {createMedicalRecord})
% Syntactic Similarity: 0.8333333333333334

The {\em e-document case study} is the most difficult for our algorithms.  
Both algorithms do well on 37 of the 39 input rules.  The greedy algorithm fails to discover the conditions and constraints needed for the remaining two rules, producing instead rules that identify several employees individually by enumerating their id's.  
One of these rules is challenging because its subject condition, resource condition, and constraint are each non-trivial, and the resource condition (resource.type.id in \{Invoice,SalesOffer,Contract\}) involves a three-element set.  The evolutionary algorithm does better on these rules, discovering them in most but not all object models.  The significantly higher WSC of the mined policies produced by both algorithms, relative to the simplified original rules, is due to their difficulty with these two rules.  
The relatively large gap between the syntactic similarity and rule-wise semantic similarity for the greedy algorithm on this policy is also due to its difficulty with the aforementioned rule, and reflects the fact that relatively small syntactic differences (e.g., changing one constant in one condition) can cause a relatively large change in the meaning of a rule.

For the {\em workforce management case study}, the evolutionary algorithm produces policies that have even lower WSC (about 11\% lower) than the simplified original policy, and hence do not have perfect similarity.  The greedy algorithm produces policies with somewhat higher WSC (about 10\% higher) than the simplified original policy.  The relatively large gap between the syntactic similarity and rule-wise semantic similarity for the greedy algorithm on this policy is due to the heuristic that prefers constraints over conditions.
% (even when they are semantically equivalent, i.e., when replacing a condition with a constraint does not change the meaning of the rule). 
This heuristic typically helps increase the generality of rules, but is not helpful for some rules in this policy.  The evolutionary algorithm does not incorporate this heuristic and achieves higher syntactic similarity.

% up-to-date but omitted to save space.
% the greedy algorithm achieves an average syntactic similarity of 0.81 and an average semantic similarity of 0.90, and the evolutionary algorithm achieves an average syntactic similarity of 0.92 and an average semantic similarity of 0.96 with the simplified original policy.  The mined policy's WSC is higher than the WSC of the simplified original policy.  The gap between the syntactic and semantic similarities in greedy algorithm's result is due to cases like the following.  The input rule $\langle$Employee, subject.groups.id contains \{Provisioning\}, StockRefillNotification, resource.tenant.id = PP, true, \{receive\}$\rangle$ is replaced in the mined policy with a similar rule in which the resource condition resource.tenant.id = PP is replaced with the constraint subject.employer.id = resource.tenant.id, which is semantically equivalent because an Employee in Provisioning group must have employer PP.  This replacement is due to the algorithm's general preference for constraints over conditions, discussed in Section \ref{sec:algorithm}, although in this case it does not improve generalization. Evolutionary algorithm does not suffer from the same problem since it does not use any heuristic approach, and the algorithm is able to reconstruct the input rule using the resource condition resource.tenant.id = PP most of the time.

To evaluate the benefit of our specialized genetic operators, we ran a variant of our algorithm modified to use only the two classic genetic operators (single mutation and crossover), on the two case studies, using the same 30 pseudo-random object models for each policy.  Eliminating the specialized genetic operators reduces the average policy similarity slightly for the e-document case study (from 0.90 to 0.89 for syntactic similarity, and from 0.86 to 0.84 for rule-wise semantic similarity) and significantly for the workforce management case study (from 0.92 to 0.82 for syntactic similarity, and from 0.96 to 0.79 for rule-wise semantic similarity).

% same parameter values were used for both versions of the alg.

%For e-doc policy:
%- with generic operators only:
%Average syntactic similarity: 0.889557952209275 
%Average semantic similarity: 0.8356794849775324 
%Average WSC: 343.96666666666664 
%- with specialized operators:
%Average syntactic similarity: 0.89821479899434
%Average semantic similarity: 0.863074572588726 
%Average WSC: 325.633333333333

%For e-workforce policy:
%- with generic operators only:
%Average syntactic similarity: 0.8232723020714203 
%Average semantic similarity: 0.7901389846042478
%Average WSC: 248.53333333333333
%- with specialized operators:
%Average syntactic similarity: 0.920959691885617
%Average semantic similarity: 0.962468906575241
%Average WSC: 184.466666666666

% ran experiments with increasing nGenerationsSearch from 2000 to 10000, and nGenerationsImprove from 1000 to 5000 on 3 different object models of edoc and eWorkforce. 
% average results:
%                  e-doc  workforce
%syntactic orig	   0.92	0.94
%syntactic 5*nGen	0.92	0.96
%semantic orig	   0.94	0.96
%semantic 5*nGen   0.99	0.97

%I cannot give exact comparison on running time since I ran old experiments on my PC, and new experiments on lab PC. However, the running time of new experiments is not too bad. It takes around 8 mins on average to run a eWorkforce policy model, and it takes around  56 mins on average to run a e-doc policy model.

\subsection{Performance Results}

%timings are from \policy-mining\src\RelationalPolicyMining\experiment results\performance_experiment-07-30-2018\performance_experiment_case_studies_07_30_18.xlsx
% timings in original submission (before caching optimization): policy-mining\src\RelationalPolicyMining\experiment results\performance_experiment-04-11-2018\performance_experiment_data.xlsx
\begin{figure}[tb]
  \centering
  \includegraphics[width=0.5\textwidth]{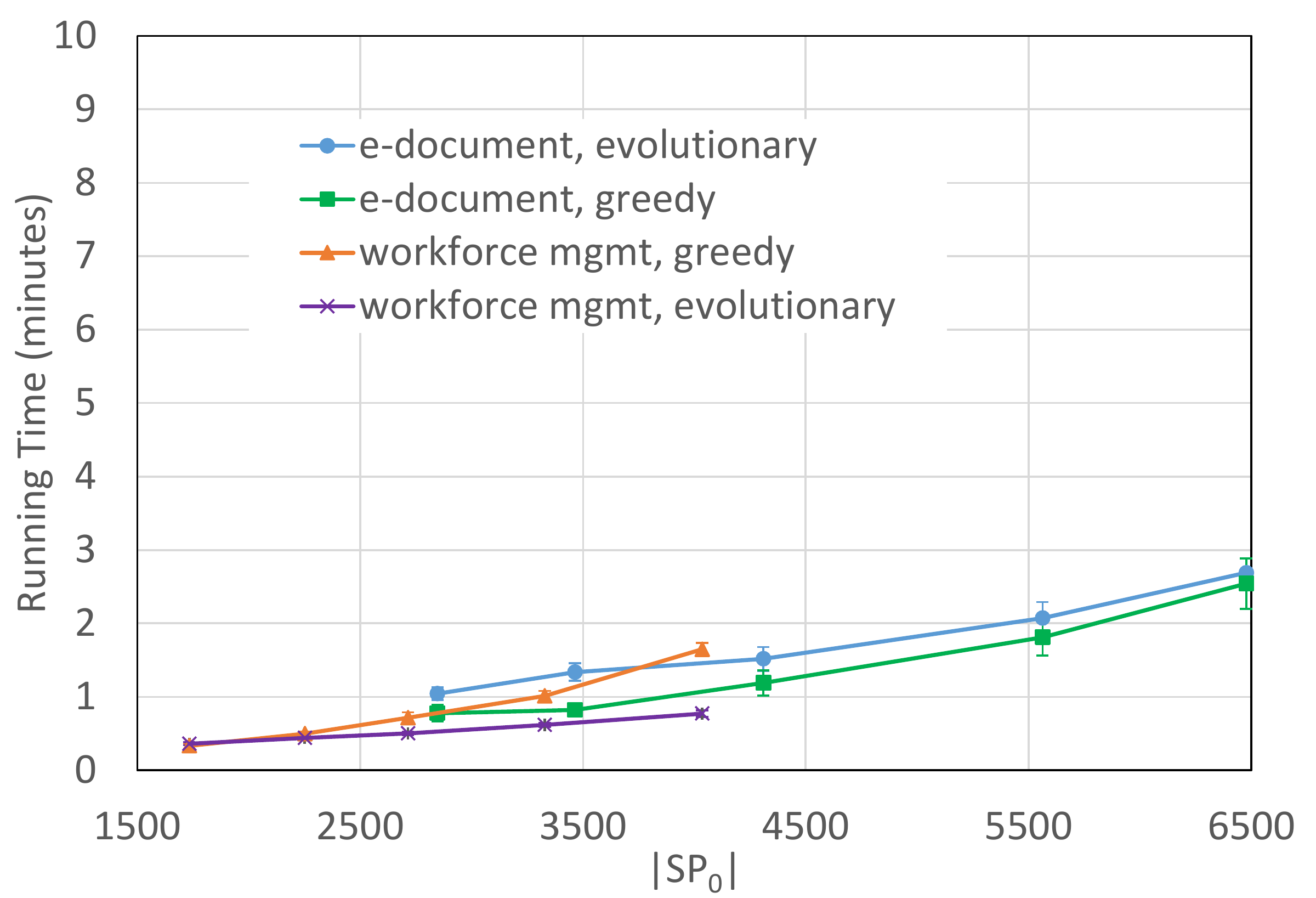}
  \caption{Running time of both algorithms on the case studies, as a function of the number of subject-permission tuples.}
  \label{fig:performance}
\end{figure}

Figure \ref{fig:performance} shows the running time of both algorithms on the case studies as a function of ACL policy size $|\spa_0|$.  Each data point is the average over 10 pseudo-random object models.  Error bars (too small to see in some cases) show 95\% confidence intervals using Student's t-distribution.  We see that the algorithms have similar performance on both case studies.  The slopes of the best-fit lines on a log-log plot of the data are: for e-document, 1.5 for the greedy algorithm, and 1.1 for the evolutionary algorithm; for workforce management, 1.9 for the greedy algorithm, and 0.9 for the evolutionary algorithm.  This is an encouraging indicator of the algorithms' scalability: they can mine dozens of complex rules from ACLs with several thousand entries in minutes, and the growth in running time, as a function of the number of ACL entries, is less than quadratic for the greedy algorithm, and is close to linear for the evolutionary algorithm.

% !TeX root = main.tex

\section{Conclusions and Future Work}

A long-standing trend in research on access control policy mining is to handle increasingly expressive policy languages, starting with flat RBAC \cite{conf/sacmat/kuhlm03}, advancing to RBAC with role hierarchy \cite{conf/sacmat/schle05}, followed by RBAC with extensions such as temporal constraints \cite{mitra13toward} and parameterized roles \cite{xu13mining}, and then ABAC \cite{xu15miningABAC}.  This paper continues that trend.  We introduced ORAL, an ReBAC policy language formulated as an object-oriented extension of ABAC, defined the ReBAC policy mining problem, and presented the first two algorithms for that problem.  Our evaluation on four sample policies and two larger and more complex case studies, based on SaaS applications offered by real companies, demonstrate the effectiveness of our algorithms.

There are many interesting directions for future work on access control policy mining.  One obvious direction is policy mining for ReBAC languages with additional features: additional data types and corresponding relational operators (e.g., integers with inequalities), negation, temporal or spatial constraints, actions involving multiple resources, etc.  Another practical direction for future work is mining ReBAC policies from incomplete data (e.g., access logs instead of ACLs) or noisy data (e.g., extraneous permissions, or incorrect attribute values).  Yet another direction is to explore incremental approaches to policy mining to support policy evolution.

\myparagraph{Acknowledgements}  We thank Eric Medvet for helpful discussions about grammatical evolution.

\svonly{
\begin{acknowledgements}
\thanksText
\end{acknowledgements}}

%\svonly{\myparagraph{Conflict of Interest} The authors declare that they have no conflict of interest.}

% svjournal's spphys bibstyle does not print titles of conference papers.  I used bibstyle plain instead, since it also uses numeric references.
% 
\acmonly{\bibliographystyle{ACM-Reference-Format}}\lncsonly{\bibliographystyle{splncs03}}\articleonly{\bibliographystyle{alpha}}\svonly{\bibliographystyle{plain}}
\bibliography{references}
%\IfFileExists{references.bib}{\bibliography{references}}{\bibliography{../../references}}

\end{document}

%%% Local Variables:
%%% mode: latex
%%% TeX-master: t
%%% TeX-PDF-mode: t
%%% End: